\documentclass[12pt]{article}
\usepackage[dvips]{graphicx}
\newcommand{\eps}{\varepsilon}
\newcommand{\omegad}{\widetilde{\omega}}
\newcommand{\rb}{{\bf r}}
\newcommand{\Hp}{{\cal H}_p}
\newcommand{\Hd}{\overline{H}}
\newcommand{\mod}{\mathop{\rm mod}\nolimits}
\newcommand{\arctg}{\mathop{\rm arctg}\nolimits}

\author{V.V.Sidorenko\\ {\small \it sidorenk@spp.keldysh.ru}\\
{\small \it Keldysh Institute of Applied Mathematics}
\\ {\small \it Moscow, RUSSIA}}
\title{Long-term evolution of the asteroid orbits \\
at the 3:1 mean motion resonance with Jupiter (planar problem)}

\date{}
\begin{document}
\maketitle
{\small
\begin{quotation}
The 3:1 mean-motion resonance of the planar elliptic restricted
three body problem (Sun-Jupiter-asteroid) is considered. The
double numeric averaging is used to obtain the evolutionary
equations which describe the long-term behavior of the asteroid's
argument of pericentre and eccentricity. The existence of the
adiabatic chaos area in the system's phase space is shown.
\end{quotation}}

\begin{center}
{\large \bf 1. Introduction}
\end{center}

The 3:1 mean-motion resonance in the planar elliptic restricted
three-body problem (Sun--Jupiter--asteroid) has long attracted
considerable attention of specialists [1--16]. In order to find
secular effects, the equations of motion can be averaged over fast
variables, namely, over mean longitudes of the asteroid and
Jupiter (see, for example, [3, 12]). Upon averaging, a
nonintegrable system appears which describe the "fast" and "slow"
components of secular evolution. The "fast" evolution consists in
changing resonance phase (Delaunay variable)
$D=\lambda-3\lambda'$, where $\lambda$ and $\lambda'$ are the mean
longitudes of the asteroid and Jupiter, respectively. The "slow"
evolution reveals itself in a gradual change of perihelion
longitudes of resonance asteroid orbits.

In [4--6,8,11] a model Hamiltonian system was considered with
Hamiltonian $\Hp$ which was a principal part of the Hamiltonian
of a planar elliptic three-body problem averaged over $\lambda$
and $\lambda'$ taking the resonance into account. Such a
truncation of the Hamiltonian is justified at small orbit
eccentricities of the asteroid and Jupiter. Processes described
by this model system are also separated into "fast" and "slow"
processes. In order to analyze different variants of the "slow"
evolution, one can make yet another averaging: averaging over
fast processes [6,17].

In this paper double averaging is used for studying the "slow"
evolution without restrictions on the orbit eccentricity of an
asteroid. The results of this study allow one to understand,
under which initial conditions the system with Hamiltonian $\Hp$
correctly describes secular effects at the resonance under
discussion.
\newpage

\begin{center}
{\large \bf 2. Averaging over mean longitudes}
\end{center}

We assume that the semimajor axis of the orbit of Jupiter can be
taken as the unit length, while the sum of masses of the Sun and
Jupiter is the unit mass. The unit time is chosen so that the
period of revolution of Jupiter around the Sun is equal to $2\pi$.

We write the equations of motion of the asteroid in the variables
$$
x,\;y,\;L,\;D,
$$
where $x,y,$ and $L$ are the elements of the second canonical
Poincare system, and they are related to osculating elements by
the formulas
$$
x= \sqrt{2\sqrt{(1-\mu)a}[1-\sqrt{1-e^2}]}\cos\omegad,\;
\eqno(2.1)
$$
$$
y=-\sqrt{2\sqrt{(1-\mu)a}[1-\sqrt{1-e^2}]}\sin\omegad,\;
$$
$$
L = \sqrt{(1-\mu)a}.
$$
Here, $\omegad, e,$ and $a$ are the longitude of perihelion,
eccentricity, and semimajor axis of the asteroid orbit, and $\mu$
is the mass of Jupiter.

The equations of motion have the following canonical form
$$
\frac{dx}{dt} = - \frac{\partial \cal K}{\partial y}\,,\quad
\frac{dy}{dt} =  \frac{\partial \cal K}{\partial x}\,,\;
\eqno(2.2)
$$
$$
\frac{dL}{dt} = - \frac{\partial \cal K}{\partial D}\,,\quad
\frac{dD}{dt} =  \frac{\partial \cal K}{\partial L}
$$
with the Hamiltonian
$$
{\cal K} = - \frac{(1-\mu)^2}{2L^2} - 3L - \mu R. \eqno(2.3)
$$
Function $R$ in the expression for $\cal K$ is defined in the
following way:
$$
R=\frac{1}{\left|\rb - \rb'\right|} - \frac{(\rb,\rb')}{{r'}^3}\,,
$$
where $\rb=\rb(x,y,L,\lambda(D,\lambda'))$ and
$\rb'=\rb'(\lambda')$ are the heliocentric radii vectors of the
asteroid and Jupiter.

Formal averaging of the equations of motion consists in
substitution of the function
$$
W(x,y,L,D)= \frac{1}{2\pi} \int_0^{2\pi}
R(x,y,L,D,\lambda')\,d\lambda'. \eqno(2.4)
$$
for function $R$ in expression (2.3) for Hamiltonian.

As a result, the equations of motion become autonomous: the mean
longitude of Jupiter $\lambda'=t+\lambda'_0$ is eliminated from
their right-hand sides.

A detailed description of the numerical algorithm used to
construct averaged equations at resonance 3:1 in the general case
(the motion of the asteroid is not restricted to the orbit of
Jupiter) is given in [18,19]. In the planar problem the similar
formulas can be written in less cumbersome way (see Appendix).
\newpage

\begin{center}
{\large \bf 3. Fast-slow system describing secular effects in the
motion of a resonance asteroid}
\end{center}

In motion of the asteroid in resonance 3:1 with Jupiter the value
of variable $L$ is close to $L_0= 1/\sqrt[3]{3}$. In the limiting
case $\mu=0$ the asteroid, moving along the orbit with the
semimajor axis $a_{res}=L_0^2\approx 0.48074$, makes exactly
three revolutions around the Sun during one revolution of Jupiter.

Following the general scheme of studying resonance effects in
Hamiltonian systems [20] we change variable $L$ in system (2.2)
averaged over $\lambda'$ for variable $\Phi=(L_0-L)/\sqrt{\mu}$
representing the normalized deviation of $L$ from its resonance
value and introduce a new independent variable $\tau =
\sqrt{\mu}t$. Restricting ourselves to the leading terms in the
expansion in terms of $\eps=\sqrt{\mu}$ of the right-hand sides
of the equations of motion in variables $x,y,\Phi,$ and $D$ we
get:
$$
\frac{dx}{d\tau}=\eps\frac{\partial V}{\partial y}\,,\;
\frac{dy}{d\tau}=-\eps\frac{\partial V}{\partial x}\,, \eqno(3.1)
$$
$$
\frac{dD}{d\tau}=\alpha\Phi\,,\;
\frac{d\Phi}{d\tau}=-\frac{\partial V}{\partial D}\,,
$$
where
$$
V(x,y,D)=W(x,y,L_0,D)\,,\; \alpha=\frac{3}{L_0^4}=9\sqrt[3]{3}\,.
$$

If we take as conjugate canonical variables $x/\sqrt{\eps}$ and
$y/\sqrt{\eps}$, $D$ and $\Phi$, system (3.1) is Hamiltonian with
the Hamilton function
$$
{\cal H} = \frac{\alpha \Phi^2}{2} + V(x,y,D).
$$

In the general case, variables $x,y,D,$ and $\Phi$ have different
rates of variation:
$$
\frac{dD}{d\tau},\frac{d\Phi}{d\tau}\sim 1,\;
\frac{dx}{d\tau},\frac{dy}{d\tau}\sim \eps.
$$
We will refer to variables $D,\Phi$ and $x,y$ as "fast" and "slow"
variables, respectively.

System (3.1) allows one to investigate secular effects in the
dynamics of resonance asteroids without constraints on the orbit
eccentricity value. Evolution of osculating elements $e$ and
$\omegad$ with an error $O(\eps)$ is described by the relations
$$
e=\frac{1}{2L_0}\sqrt{(x^2+y^2)\left[4L_0 - (x^2+y^2)\right]}\,,
\eqno(3.2)
$$
$$
\omegad=\left\{
\begin{array}{cc}
2\pi - \arccos \frac{\displaystyle x}{\displaystyle \sqrt{x^2 +
y^2}}, & y \geq 0
\\
\rule{0pt}{8mm} \arccos \frac{\displaystyle x}{\displaystyle
\sqrt{x^2 + y^2}}, & y < 0
\end{array} \right.
(x^2+y^2 \ne 0).
$$
where variables $x$ and $y$ change in the
region$S=\{(x,y),x^2+y^2<2L_0\}$.

Taking into account the separation of variables into "fast" and
"slow" variables, we refer to system (3.1) as a fast-slow (FS)
system [20].\vspace*{2.5cm}

\begin{center}
{\large \bf 4. Principal part $\Hp$ of FS-system Hamiltonian}
\end{center}

In [4, 6, 8, 17] the properties of a Hamiltonian system with
Hamiltonian $\Hp$ were studied. After changing some designations
it takes on the form:
$$
\Hp = \frac{\alpha \Phi^2}{2} + V_p(x,y,D). \eqno(4.1)
$$
Here
$$
V_p(x,y,D)=
{e'}^2\overline{V}\left(\frac{x}{e'},\frac{y}{e'},D\right)\!,
\eqno(4.2)
$$
$$
\overline{V}=\lim_{e'\to 0} e'^{-2} \left(
V(e'x,e'y,D)-V_0\right)-V_2.
$$
Constants $V_0$ and $V_2$ in (4.2) have the following values:
$$
V_0 = \frac{b_1^{(0)}(a_{res})}{2}=
\frac{2}{\pi}K(a_{res})\approx 1.0667,
$$
$$
V_2 = \left.\frac{d}{da}\left( \frac{a^2}{2} \frac{db_1^{(0)}}{da}
\right)\right|_{a=a_{res}}\approx 0.1422.
$$
In the above formulas $b_1^{(0)}(a_{res})$ is the Laplace
coefficient, and $K(a_{res})$ is the complete elliptic integral of
the first kind with modulus $a_{res}$.

Let us assume that $(x,y)\in S_W=\{(x,y),x^2+y^2<C_W^2 e'^2\}$,
where $C_W \sim 1$. In this case the following estimations are
valid:
$$
\left|\frac{\partial V}{\partial D}\right|\sim
\left|\frac{\partial V_p}{\partial D}\right|\sim e'^2,\;
\left|\frac{\partial V}{\partial D}- \frac{\partial V}{\partial
D}\right|\sim e'^4, \eqno(4.3)
$$
$$
\left|\frac{\partial V}{\partial \xi}\right|\sim
\left|\frac{\partial V_p}{\partial \xi}\right|\sim e',\;
\left|\frac{\partial V}{\partial \xi}- \frac{\partial
V_p}{\partial \xi}\right|\sim e'^3,\;\xi=x,y.
$$
Taking advantage of relationships (4.3), one can demonstrate
that, when independent variable $\tau$ changes in the interval
$\sim \eps^{-1}$, the behavior of fast and slow variables in the
solution to FS-system (3.1) which satisfies the condition
$(x(\tau),y(\tau))\in S_W$ is described by analogous components
of the solution with Hamiltonian (4.1) at the same initial data
with accuracies of $~\eps^{-1}{e'}^4$ and $~{e'}^3$, respectively.

Investigation of the system with Hamiltonian (4.1) made by J.
Wisdom can be interpreted as an analysis of the properties of
solutions with slow variables in $S_W$ in the case when $C_W
\approx 5$. Assuming in what follows $C_W=5$, we will refer to
$S_W$ as the Wisdom region.

There is an important difference in the behaviors of $V(x,y,D)$
and $V_p(x,y,D)$ which should be taken into account in future. The
Fourier series of function $V(x,y,D)$ includes infinite number of
harmonics at any $(x,y)\in S$ (point $(0,0)$ can be an exception:
if eccentricity of Jupiter $e'=0$, then $V(0,0,D)\equiv V_0$).
The expansion of $V_p(x,y,D)$ into a Fourier series in terms of
$D$ contains a finite number of harmonics [4]:
$$
V_p(x,y,D)=B(x,y)+A(x,y)\cos(D-R(x,y)). \eqno(4.4)
$$
At points
$$
P_1=(x_1,0),\; x_1= e'\bar{x}_1,\;\bar{x}_1 \approx 0.14341
$$
and
$$
P_2=(x_2,0),\; x_2= e'\bar{x}_2,\;\bar{x}_2 \approx 2.93415
$$
the value of function $A(x,y)$ is equal to zero.

\begin{center}
{\large \bf 5. Properties of the fast subsystem}
\end{center}

At $\eps=0$ the equations of fast variables coincide with the
equations of motion of a Hamiltonian system with a single degree
of freedom which include $x$ and $y$ as parameters:
$$
\frac{dD}{d\tau}=\alpha\Phi\,,\;
\frac{d\Phi}{d\tau}=-\frac{\partial V}{\partial D}\,. \eqno(5.1)
$$
The qualitative behavior of trajectories on the phase portrait of
system (5.1) is determined by the properties of function
$V(x,y,D)$.

It turns out that there are important distinctions in fast
dynamics at $e'>e_*'$ and at $e'<e_*'$, where $e_*'\approx
0.0385$ is the minimum eccentricity of the Jupiter orbit
admitting its intersection with the orbit of the resonance
asteroid at $\mu=0$.

In the case $e'<e_*'$ (orbits of resonance asteroids do not
intersect the Jupiter orbit at any $e < 1$) function $V(x,y,D)$ is
limited. If $(x,y) \notin S_1 \bigcup S_2$, where $S_1$ and $S_2$
are certain $O(e'^3)$-neighborhoods of points $P_1$ and $P_2$,
then the phase portrait of fast subsystem (5.1) is topologically
equivalent to the phase portrait of a mathematical pendulum (Fig.
1a). In what follows we will designate as $S_0$ the set of $(x,y)$
values at which dynamics in the fast subsystem is of "pendulum"
type.

The coefficient of the first harmonic in the Fourier expansion of
potential $V(x,y,D)$ vanishes at points $P_1$ and $P_2$.
Therefore, when $(x,y)$ are selected in the neighborhood of these
points the qualitative behavior of $V(x,y,D)$ is determined by the
next harmonics of the Fourier expansion. At $(x,y) \in S_1 \bigcup
S_2$ potential $V(x,y,D)$ is characterized (in the general case)
by the existence of local minimum and maximum, which leads to a
more complicated pattern of motion in the fast subsystem (Fig.
1b). The boundaries of regions $S_1$ and $S_2$ are curves close
to astroid, a typical situation taking place when reconstructions
of the potential depending on two parameters are studied [20].

In the case $e'>e_*'$, at appropriate choice of the initial value
of variable $D$, the motion of the asteroid along the orbit
crossing the Jupiter orbit will be accompanied by formal
collisions resulting in divergence of the integrals determining
functions $V(x,y,D)$. Let us designate the set of $(x,y)$ values
corresponding to the resonance orbits with intersections of the
orbit of Jupiter as $S_3$. Figure 1c demonstrates an approximate
pattern of the phase portrait of the fast subsystem at $(x,y)\in
S_3$. The partition of $S$ into regions with different types of
fast dynamics is shown in Fig. 2.

Let
$$
\Phi(\tau,x,y,h)\,,\; D(\tau,x,y,h) \eqno(5.2)
$$
be a solution to Eq. (5.1) satisfying the condition
$$
{\cal H}(x,y,\Phi(\tau,x,y,h),D(\tau,x,y,h))=h,
$$
in which variable $D$ changes in rotational or oscillating manner:
$$
D(\tau+T,x,y,h) = D(\tau,x,y,h)\mod 2\pi
$$
Here, $T(x,y,h)$ is the period of the solution. We associate this
solution with the following quantity
$$
I(x,y,h)=\frac{\alpha}{2\pi} \int_0^{T/\sigma}
\Phi^2(\tau,x,y,h)\,d\tau,
$$
where the value of $\sigma$ is determined by the type of
solution. For rotational solutions $\sigma=1$ and, hence,
$I(x,y,h)$ is the action integral. For oscillating solutions
$\sigma=2$, and the value of $I(x,y,h)$ equals a half of the
action integral.

At $\eps\ne 0$ variables $x(\tau)$ and $y(\tau)$ in the right-hand
sides of Eqs. (5.1) can be interpreted as slowly varying
parameters. The quantity $I(x,y,h)$, coinciding to a constant
factor with the action integral, will be an adiabatic invariant
of system (3.1).

\begin{center}
{\large \bf 6. Averaging along solutions to the "fast" subsystem}
\end{center}

Averaging the right-hand sides of the equations for $x,y$ in
system (3.1) along solutions to fast subsystem (5.2) we get
evolution equations describing to the error $O(\eps)$ the changes
of variable $x,y$ on the time interval with duration $\sim
1/\eps$ (or $\sim 1/\mu$ in terms of original time units):
$$
\frac{dx}{d\tau}=\eps\left\langle \frac{\partial V}{\partial
y}\right\rangle\,,\; \frac{dy}{d\tau}=-\eps\left\langle
\frac{\partial V}{\partial y}\right\rangle\,. \eqno(6.1)
$$
Here
$$
\left\langle\frac{\partial V}{\partial \zeta}\right\rangle =
\frac{1}{T(x,y,h)} \int_0^{T(x,y,h)} \frac{\partial
V}{\partial\zeta} (x,y,D(\tau,x,y,h),\Phi(\tau,x,y,h))\,d\tau\,
$$
$$
\zeta=x,y.
$$

Construction of phase portraits of system (6.1) is the efficient
method of studying the evolution of slow variables $x,y$. At
different $h$ the phase portraits can differ in the number of
equilibrium positions and in the behavior of separatrices. In
sections {\bf 9,10} we present numerous examples allowing one to
get a general idea about the character of reconstructions of the
phase portrait when $h$ changes.

\begin{center}
{\large \bf 7. Auxiliary functions}
\end{center}

Following [17], let us consider the functions
$$
H_*(x,y)= \min_D V(x,y,D)\,,\; H^*(x,y)= \max_D V(x,y,D).
\eqno(7.1)
$$
The behavior of functions $H_*(x,y)$ and $H^*(x,y)$ in $S$ is
determined by the eccentricity $e'$ of the orbit of Jupiter.

At $e'=0$, the values of the functions at point $(x,y)$ depend
exclusively on its distance from the origin of coordinates:
$$
H_*(x,y)=H_*(r),\;H^*(x,y)=H^*(r),
$$
where $r=\sqrt{x^2+y^2}$. The plots of functions $H_*(x,y)$ and
$H^*(x,y)$ in the case $e'=0$ are presented in Fig. 3a. At the
origin of coordinates $x=y=0$ functions $H_*$ and $H^*$ have the
same value:
$$
H_*(0,0)=H^*(0,0)=V_0.
$$
This value is the global maximum and global minimum for functions
$H_*$ and $H^*$, respectively. The minimum value of functions
$H_*$ is reached at the distance $r_*\approx 0.74420$ from the
origin of coordinates. Function $H^*$ reaches its maximum value
at the distance $r^*\approx 1.05431$.

The plots of functions $H_*$ and $H^*$ at $e'\in(0,e_*')$ can be
considered as a result of deformation of the plots constructed
for the case $e'=0$. The periphery parts of the $H_*$ plot are
elevated in the half-plane $x>0$ and go down in the
half-plane$x<0$ (Fig. 3c). Function $H_*$ have a global minimum at
point $P_*=(-r_*+O(e'),0)$, and its saddle point is located at
$P_{**}=(r_*+O(e'),0)$. The sides of the $H^*$ plot are displaced
in opposite directions: point $P^*=(-r^*+O(e'),0)$ becomes a
point of the global maximum, and the saddle is located at point
$P^{**}=(r^*+O(e'),0)$. At $e'\to e_*'$ the point $P^*$ is
shifted to the boundary of the region, and the value of $H^*$ at
this point increases infinitely.

In the Wisdom region $S_W$
$$
H_*(x,y)=V_0 + {e'}^2 \left[
\Hd_*\left(\frac{x}{e'},\frac{y}{e'}\right) + V_2 \right] +
O({e'}^4), \eqno(7.2)
$$
$$
H^*(x,y)=V_0 + {e'}^2 \left[
\Hd^*\left(\frac{x}{e'},\frac{y}{e'}\right) + V_2 \right] +
O({e'}^4),
$$
where
$$
\Hd_*= \min_D \overline{V}(x,y,D),\;\Hd^*= \max_D
\overline{V}(x,y,D).
$$
Taking advantage of the results of studying the properties of
functions $\Hd_*$ and $\Hd^*$ presented in [17], one can
establish that function $H_*$ has the global maximum at a certain
point $P'_2 \in S_2$, a local minimum at point $P'_1 \in S_1$,
and a saddle at the point
$$
P_H =(x_H,0),\;x_H=e'\bar{x}_H+o(e'),\;\bar{x}_H\approx 1.33646.
$$

Function $H^*$ will have the global minimum at point $P''_1\in
S_1$, a local maximum at point $P''_2 \in S_2$, and a saddle at
the point
$$
P_S=(x_S,0),\; x_S=e'\bar{x}_S+o(e'),\; \bar{x}_S\approx 1.86718.
$$

We will designate the minimum (maximum) values of functions $H_*$
and $H^*$ in $S$ as $h_{*min}$ and $h_{min}^*$ ($h_{*max}$ and
$h_{max}^*$), respectively. The value of function $H_*$ at point
$P_*$ is $h_{**}$, and the value of $H^*$ at point $P^*$ is
$h^{**}$.

At $e'>e'_*$ the behavior of $H^*(x,y)$ changes substantially:
$H^*(x,y)=\infty$ for $(x,y)\in S_3$ (Fig. 3c).

\begin{center}
{\large \bf 8. Forbidden area and uncertainty curve on phase
portraits}
\end{center}

The region
$$
M(h)=\{(x,y)\in S, H_*(x,y)>h\} \eqno(8.1)
$$
is forbidden for phase trajectories of system (6.1). At a given
$h$ the slow variables cannot assume values from $M(h)$.

The curve
$$
\Gamma(h)=\{(x,y)\in S, H^*(x,y)=h\}
$$
is called the uncertainty curve. In the case $H^*(x,y)=h$ the
trajectory of the fast system is a separatrix and, hence, one
cannot use averaging. If $\Gamma(h)$ is present on the phase
portrait of system (6.1), it consists of several fragments
undergoing a series of bifurcations when $h$ is varied.

When a projection of the phase trajectory of system (6.1) onto
the plane $x,y$ crosses the curve $\Gamma(h)$, a quasi-random
jump of adiabatic invariant $I(x,y,h)$ occurs [7]. When studying
the evolution of slow variables on a time interval of order of
$1/\eps$, this violation of adiabatic invariance is usually
neglected, and solutions of the averaged system on curve
$\Gamma(h)$ are glued in accordance with the following rule
$$
I_{before}=I_{after},
$$
where $I_{before}$ is the value of $I(x,y,h)$ along the part of
the phase trajectory of system (6.1) approaching $\Gamma(h)$, and
$I_{before}$ is the value of $I(x,y,h)$ on the trajectory part
going away from curve $\Gamma(h)$. For most initial conditions,
the accuracy of such an approximation is $O(\eps)$ on the
specified time interval.

The phenomena taking place at multiple intersections of the
uncertainty curve will be discussed in Sec.{\bf 13}.

\begin{center}
{\large \bf 9. Studying slow evolution based on averaged
equations: the case $e'<e_*'$}
\end{center}

Figures 4--11 present examples of phase portraits constructed with
the help of numerical integration of Eqs. (6.1) under an
assumption that Jupiter moves along the orbit with eccentricity
$e'=0.02 < e_*'$.

For better visualization the phase portraits present the behavior
of quantities $\widehat{x}$ and $\widehat{y}$ which are related to
variables $x,y$ and osculating orbital elements $e, \omegad$ as
$$
\widehat{x}=\frac{x}{2L_0}\sqrt{4L_0 - (x^2+y^2)}= e\cos\omegad,
$$
$$
\widehat{y}=\frac{y}{2L_0}\sqrt{4L_0 - (x^2+y^2)}=-e\cos\omegad.
$$
In the case $h \in (h_{*min},h_{**})$ system (6.1) has one stable
position of equilibrium (Fig. 4). Phase trajectories are located
in the region $S/M(h)$ which shrinks to point $P_*$ at $h \to
h_{*min}$.

Bifurcation at $h=h_{**}$ leads to the appearance on the phase
portrait of one more position of equilibrium (Fig. 5). The
forbidden area $M(h)$ is split into two components (central and
periphery zones). With increasing $h$ the periphery zone is
gradually contracted and disappears.

In the interval $I_W=(h^*_{min} - C_W{e'}^2,h_{*max} +
C_W{e'}^2)$ a series of bifurcations occurs in which changing
behavior of phase trajectories in $S_W$, and of geometry of
$\Gamma(h)$ and $M(h)$ approximately corresponds to a scenario
described in [17]. Insignificant distinctions are caused by the
difference (discussed in Sec. {\bf 4}) in the properties of
function $V(x,y)$ and of $V(x,y)$ in the neighborhood of points
$P_1$ and $P_2$. They reveal themselves in two very narrow
intervals of $h$ values ($\sim e'^4$) whose lower boundaries are
minimum values of $H_*$ on the boundaries of regions $S_1$ and
$S_2$, respectively. The approximate formula
$$
h = V_0 + e'^2(\overline{h} - V_2) \eqno(9.1)
$$
allows one to relate bifurcation values of parameter $h \in I_W$
with bifurcation values of parameter $\overline{h}$ of the Wisdom
system calculated in [17].

As an illustration, Fig. 6a presents the phase portrait of system
(6.1) at $h=1.068 \in I_W$. Figure 6b gives its enlarged fragment
that demonstrates the behavior of phase trajectories in the Wisdom
region. One can notice a similarity of Fig. 6b with Fig. 5 of [6]
and Fig. 11 of [17] where phase portraits of the Wisdom system
are presented.

At $h\approx 1.119$ a reconnection of separatrices takes place
(Figs. 7--9). At $h\approx 1.151$, the pair of equilibrium
positions located on the right of the $Oy$ axis (Fig. 10)
disappears from the phase portrait. At $h\approx 1.170$, the pair
of equilibria to the left of this axis merges. After that, a
single stable position of equilibrium remains in the central part
of the $S$ region (Fig. 11).

Further increase of $h$ results in the appearance of equilibrium
positions of the periphery of $S$ region (Fig. 12). Changing
geometry of the uncertainty curve at $h=h^{**}$ and its
disappearance from the phase portrait at $h=h^{*}_{max}$ are
accompanied by merging of the positions of equilibrium at points
$P^{**}$ and $P^*$, respectively.

\begin{center}
{\large \bf 10. Investigation of slow evolution based on averaged
equations: the case $e'>e_*'$}
\end{center}

If $e'>e_*'$, one needs to take into account the existence of
resonance orbits crossing the orbit of Jupiter (the orbits with
parameters from region $S_3$). Region $S_3$ in variables
$\widehat{x},\widehat{y}$ looks like a narrow belt on the
periphery of region $S$. If one chooses $(x,y)\in S_3$ in system
(3.1), there are, in the general case, two "fast" processes over
which averaging is possible. Therefore, the right-hand sides of
Eqs. (6.1) are ambiguously determined in $S_3$. At numerical
integration, in the situation when evolving orbit of a resonance
asteroid begins to intersect the orbit of Jupiter, one can choose
an appropriate solution to fast subsystem (5.1) as a closest to
the solution along which averaging was made on the preceding step.

Figures 13-16 present the phase portraits of system (6.1)
constructed for the case $e'=0.048$. To choose such value of
eccentricity of Jupiter is traditional for numerical
investigations of the dynamics of asteroids in the context of the
restricted elliptical three-body problem [4].

The magnified fragments of the phase portrait at $h=1.07146$
(Figs. 13b and 13c) resemble the phase portraits presented on
Figs. 9 and 10 of paper [17]. These figures show that system
(6.1) has the families of closely located equilibrium positions.

Figures 14--16 demonstrate the reconnection of separatrices at $h
\approx 1.0955$. They resemble Figs. 7--9 illustrating a similar
bifurcation in the case when Jupiter moves along the orbit with
eccentricity $e'=0.02$. Nevertheless, there are important
qualitative distinctions: uncertainty curve $\Gamma(h)$ intersects
seperatrices in Figs. 14--16, while there is nothing of this kind
in Figs. 7--9. The location of the uncertainty curve on the phase
portrait determines the size of the region of adiabatic chaos in
the phase space of a nonaveraged system (Sec. {\bf
13}).\vspace*{1.5cm}

\begin{center}
{\large \bf 11. Bifurcation diagrams}
\end{center}

Numerical integration of Eqs. (6.1) shows that in most cases
qualitative changes in the behavior of trajectories on phase
portraits are caused by some changes in properties of the
equilibrium solutions lying on the $Ox$ axis. Therefore, when
studying the bifurcations of phase portraits, of fundamental
importance is the analysis of diagrams that demonstrate the types
and position of equilibrium solutions on the $Ox$ axis at
different values of $h$.

As an example, Figs. 17a and 18a present the bifurcation diagrams
constructed under an assumption that the eccentricity of Jupiter
$e'$ was equal to $0.02$ and $0.048$, respectively. The left parts
of these diagrams contain the families of stable equilibrium
solutions generated for $h=h_{*min}$ at point $P_*$ and the
families of unstable equilibrium solutions generated for
$h=h_{**}$ at point $P_{**}$. With increasing $h$ these
equilibrium solutions are merged with equilibrium solution of two
other families originating in the central part of $S$. The
right-hand parts of the diagrams indicate to merging of
equilibrium solutions that takes place at critical points of
function $H^*(x,y)$. For clearness Figs. 17b and 17c present
enlarged fragments of the bifurcation diagram in the case
$e'=0.02$. In the case $e'=0.48$ there is only one similar
bifurcation in the right-hand part of the diagram: at $h=h^{**}$.

The segments of bifurcation diagrams describing reconstructions
of phase portraits in the Wisdom region are practically merged in
Figs. 17a and 18a when $h$ varies in the interval $I_W$. A blowup
image of corresponding fragment of the diagram for $e'=0.048$ can
be seen in Fig. 18b. If one compares Fig. 18b with a bifurcation
diagram of the Wisdom system presented in [17], some distinctions
(along with doubtless similarity) should be noted. In particular,
Fig. 18b indicates that the stability of equilibrium solutions in
the lowest family changes at $h\approx 1.0716$, i.e., when the
values of function $H^*$ at point $P_s$ are exceeded
insignificantly $(H^*(x_S,0)\approx 1.0715)$. In the Wisdom
system, a similar bifurcation occurs at the parameter value
substantially different from the value of auxiliary function at
the saddle point. It is quite likely that the differences are due
to the presence of singular points of function $V$ at $e'>e'_*$.
In their neighborhood the value of this function increases
infinitely. The singularities can substantially reduce the size of
the region where the function's behavior can be studied based on
consideration of its principal part.

Further detailed elaboration of the bifurcation diagram
description is related to the analysis of reconstructions of
phase portraits in the $S_1$ region at $h \in I_1 =
(H_*(x_1,0),h^*_{min}+O(e'^4))$ and in $S_2$ at $h \in I_2 =
(h_{*max},H^*(x_2,0)+O(e'^4))$. It is more complicated to seek
equilibrium solutions in these cases: as in the $S_3$ region, the
right-hand sides of averaged equations can be determined
ambiguously. We think that such a situation in the fast-slow
system of form (3.1) deserves special consideration. Therefore,
we do not present here the results of performed numerical
investigation. Since intervals $I_1$ and $I_2$ are very small
($\sim e'^4$), the absence of information on bifurcations in
$S_1$ and $S_2$ will be insignificant for a specialist who is
going to use the results of this work when studying the dynamics
of resonance asteroids. Even in the enlarged fragment of the
bifurcation diagram in Fig. 18b one cannot see any families of
equilibrium solutions in $S_1$ and $S_2$ at $h\in I_1$ and $h\in
I_2$, respectively.

\begin{center}
{\large \bf 12. Periodic solutions and invariant tori of the
three-body problem at the 3:1 mean motion resonance}
\end{center}

In the general case, the stable (unstable) periodic solutions to
FS-system (3.1) and stable (unstable) invariant tori in the
extended phase space $x,y,L,D,\lambda'$ of original system (2.2)
correspond to the stable (unstable) stationary solutions to
evolution equations (6.1).

Taking advantage of the fact that points
$$
P_*=(x_*,0),\;P_{**}=(x_{**},0),\;P_S=(x_S,0)
$$
and
$$
P^*=(x^*,0),\;P^{**}=(x^{**},0),\;P_H=(x_H,0)
$$
are critical points of functions $H_*(x,y)$ and $H^*(x,y)$,
respectively, one can prove the existence of the following
stationary solutions to FS-system (3.1)
$$
x\equiv x_*,\; y\equiv 0,\; D\equiv \pi,\; \Phi\equiv 0\;
(h=h_{*min}), \eqno(12.1)
$$
$$
x\equiv x_{**},\; y\equiv 0,\; D\equiv \pi,\; \Phi\equiv 0\;
(h=h_{**}), \eqno(12.2)
$$
$$
x\equiv x_H,\; y\equiv 0,\; D\equiv 0,\; \Phi\equiv 0\; (h=h_H),
\eqno(12.3)
$$
$$
x\equiv x^*,\; y\equiv 0,\; D\equiv 0,\; \Phi\equiv 0\;
(h=h^*_{max}), \eqno(12.4)
$$
$$
x\equiv x^{**},\; y\equiv 0,\; D\equiv 0,\; \Phi\equiv 0\;
(h=h^{**}), \eqno(12.5)
$$
$$
x\equiv x_S,\; y\equiv 0,\; D\equiv \pi,\; \Phi\equiv 0\;
(h=h_S), \eqno(12.6)
$$
in which the values of variables $x,y$ form a stationary solution
to evolution equation (6.1).

Periodic solutions of the restricted planar three-body problem
correspond to stationary solutions (12.1)--(12.6). Solutions
(12.3) and (12.6) are similar to periodic solutions studied by
Hill [1] and Sinclair [2]: in these solutions an asteroid moves
in an nearly circular orbit with eccentricity $e \sim e'$. In the
remaining cases the asteroid motion proceeds in orbits with
eccentricity $e\approx 1$ (Fig. 19). The properties of such
orbits were examined in [10, 13, 14].

\begin{center}
{\large \bf 13. Adiabatic chaos}
\end{center}

In the neighborhood of uncertainty curve $\Gamma(h)$ the
projection of a phase point of system (3.1) onto the plane $x,y$
jumps from one trajectory of averaged system (6.1) to another in a
quasi-random way: $|I_{after} - I_{before}| \approx \eps $. As a
result of a series of such jumps, the phase trajectories of
system (3.1) with close initial data
$$
|D_1(0)- D_2(0)|\approx \eps,\; |\Phi_1(0) - \Phi_2(0)|\approx
\eps,\; |x_1(0)- x_2(0)|\approx \eps,\; |y_1(0) - y_2(0)|\approx
\eps,
$$
can go away to a distance of $\sim 1$. Their projections onto the
plane $x,y$ will fill the region $\Sigma(h)$ which is a set of all
trajectories of evolution equations (6.1) intersecting
$\Gamma(h)$ (Fig. 20). In the phase space of FS-system (3.1)
diverging trajectories will be located in the region
$$
\Sigma^*(h)= \left\{ x,y,D,\Phi: {\cal
H}(x,y,D,\Phi)=h,\;(x,y)\in \Sigma(h)\right\}.
$$

We call $\Sigma^*(h)$ the region of adiabatic chaos: the complex
behavior of trajectories in $\Sigma^*(h)$ is associated with
violation of adiabatic invariance in the neighborhood of
$\Gamma(h)$.

The properties of adiabatic chaos in Hamiltonian systems were
studied in [17, 21]. The existence in this region of numerous
$(\sim 1/\eps)$ stable periodic solutions was proved in [21]. In
[17], using numerical methods, such solutions were sought for an
autonomous FS-system with two degrees of freedom.

The diverging trajectories of original (unaveraged) system (2.2)
correspond to the trajectories of system (3.1) diverging in
$\Sigma(h)$. Thus, the regions of chaotic dynamics originating
due to violations of adiabatic invariance will also exist in the
phase space of system (2.2). The stable periodic solutions to
FS-system (3.1) turn into stable invariant tori in the extended
phase space $x,y,L,D,\lambda'$.

\begin{center}
{\large \bf Conclusions}
\end{center}

Studies of 3:1 mean-motions resonance are of great importance for
understanding the evolution of orbits of many celestial bodies.
Asteroids of the Hestia family move in resonance 3:1 with Jupiter.
In [22], the possibility of such a resonance was considered for
Uranus' moons Miranda and Umbriel. A hypothesis of planet motion
in resonance 3 : 1 in the system 55 Cancri was discussed in [23].

The approach used in this paper allows us to get a sufficiently
detailed description of secular effects in motion of resonance
objects in the context of a planar restricted elliptical
three-body problem.

\begin{center}
{\large \bf Acknowledgments}
\end{center}

The author thanks M.A. Vashkoviak and A.I. Neishtadt who read the
manuscript of this paper and made useful remarks.

\newpage
\begin{center}
{\large \bf Appendix: calculation of derivatives of function $W$}
\end{center}

{\it 1. Preliminary remarks.}For calculation of $W$ it more
convenient to use instead of (2.3) the relation
$$
W = \frac{1}{6\pi} \int_0^{6\pi}
R(x,y,L,D,\lambda'(\lambda,D))\,d\lambda. \eqno(A.1)
$$
Following [18, 19], let us make in (A.1) a change for integration
over eccentric longitude $g$ equal to a sum of the eccentric
anomaly and the longitude of pericenter of an asteroid orbit
$$
W = \frac{1}{6\pi}\int_0^{6\pi} R \frac{\partial
\lambda}{\partial g}\,dg.
$$
When integrating over $g$ there is no necessity to solve the
Kepler's equation for seeking the true longitude characterizing
the position of an asteroid. The formula connecting the mean
longitude $\lambda$ with $g$ has the following form:
$$
\lambda = g - e\sin(g-\omegad). \eqno(A.2)
$$

Calculation of the derivatives of $W$ with respect to $x,y,L,$
and $D$ is reduced to differentiation of the integral over
parameter
$$
\frac{\partial W}{\partial x}= \frac{1}{6\pi}\int_0^{6\pi} \left(
\frac{\partial R}{\partial x} \frac{\partial \lambda}{\partial g}
+ R\frac{\partial^2\lambda}{\partial g
\partial x}\right)\,dg
$$
$$
\frac{\partial W}{\partial y}= \frac{1}{6\pi}\int_0^{6\pi} \left(
\frac{\partial R}{\partial y} \frac{\partial \lambda}{\partial g}
+ R\frac{\partial^2\lambda}{\partial g
\partial y}\right)\,dg
$$
$$
\frac{\partial W}{\partial L}= \frac{1}{6\pi}\int_0^{6\pi} \left(
\frac{\partial R}{\partial L} \frac{\partial \lambda}{\partial g}
+ R\frac{\partial^2\lambda}{\partial g
\partial L}\right)\,dg
$$
$$
\frac{\partial W}{\partial D}= \frac{1}{6\pi}\int_0^{6\pi} \left(
\frac{\partial R}{\partial D} \frac{\partial \lambda}{\partial
g}\right)\,dg.
$$

{\it 2.Derivatives of $R$ with respect to $x,y,L,D$.} In what
follows we use the heliocentric coordinate system
$O\xi\eta\zeta$  (all bodies move in the plane $O\xi\eta$, and the
orientation of axes is kept invariable). If projections of $\rb$
onto axes $O\xi$ and $O\eta$ are designated as $\xi$ and $\eta$
and projections of $\rb'$ as $\xi'$ and $\eta'$, then
$$
\frac{\partial R}{\partial x} = \frac{\partial R}{\partial \xi}
\frac{\partial \xi}{\partial x} + \frac{\partial R}{\partial
\eta} \frac{\partial \eta}{\partial x} + \frac{\partial
R}{\partial \xi'} \frac{\partial \xi'}{\partial x} +
\frac{\partial R}{\partial \eta'} \frac{\partial \eta'}{\partial
x}, \eqno(A.3)
$$
$$
\frac{\partial R}{\partial y} = \frac{\partial R}{\partial \xi}
\frac{\partial \xi}{\partial y} + \frac{\partial R}{\partial
\eta} \frac{\partial \eta}{\partial y} + \frac{\partial
R}{\partial \xi'} \frac{\partial \xi'}{\partial y} +
\frac{\partial R}{\partial \eta'} \frac{\partial \eta'}{\partial
y},
$$
$$
\frac{\partial R}{\partial L} = \frac{\partial R}{\partial \xi}
\frac{\partial \xi}{\partial L} + \frac{\partial R}{\partial
\eta} \frac{\partial \eta}{\partial L} + \frac{\partial
R}{\partial \xi'} \frac{\partial \xi'}{\partial L} +
\frac{\partial R}{\partial \eta'} \frac{\partial \eta'}{\partial
L},
$$
$$
\frac{\partial R}{\partial D} = \frac{\partial R}{\partial \xi'}
\frac{\partial \xi'}{\partial D} + \frac{\partial R}{\partial
\eta'} \frac{\partial \eta'}{\partial D}.
$$

{\it 3. Derivatives of $\xi,\eta$ with respect to  $x,y,L$.} One
can write the derivatives of $\xi$ with respect to $x,y,$ and $L$
that appear in (A.3) in the following form:
$$
\frac{\partial \xi}{\partial x}= \frac{\partial \xi}{\partial
e}\frac{\partial e}{\partial x} + \frac{\partial \xi}{\partial
\omegad}\frac{\partial \omegad}{\partial x}\,
$$
$$
\frac{\partial \xi}{\partial y}= \frac{\partial \xi}{\partial
e}\frac{\partial e}{\partial y} + \frac{\partial \xi}{\partial
\omegad}\frac{\partial \omegad}{\partial y}\,
$$
$$
\frac{\partial \xi}{\partial L}= \frac{\partial \xi}{\partial
a}\frac{\partial a}{\partial L} + \frac{\partial \xi}{\partial
e}\frac{\partial e}{\partial x}\,.
$$
Expressions for derivatives of $\eta$ with respect to $x,y,$ and
$L$ have a similar structure.

{\it 4. Derivatives of $\xi$ and $\eta$ with respect to Keplerian
elements.} The coordinates of an asteroid $\xi,\eta$, Keplerian
elements of its orbit $a, e, \omegad$ , and eccentric longitude
$g$ are connected by the relationships
$$
\xi=a\left[\cos g - e\cos \omegad + (1-
\sqrt{1-e^2})\sin(g-\omegad)\sin\omegad \right],
$$
$$
\eta=a\left[\sin g - e\sin \omegad - (1-
\sqrt{1-e^2)}\sin(g-\omegad)\cos\omegad \right].
$$
After simple transformations we get:
$$
\frac{\partial \xi}{\partial a}= \frac{\xi}{a}\,,\; \frac{\partial
\xi}{\partial e}= - a\left( \cos\omegad - \frac{e}{\sqrt{1-e^2}}
\sin(g-\omegad)\sin\omegad \right)\,,
$$
$$
\frac{\partial \xi}{\partial \omegad}= \left[ e\sin\omegad + (1
-\sqrt{1-e^2}) \sin(g-2\omegad)\right]\,,
$$
$$
\frac{\partial \eta}{\partial a}= \frac{\eta}{a}\,,\;
\frac{\partial \eta}{\partial e}= - a\left( \sin\omegad +
\frac{e}{\sqrt{1-e^2}} \sin(g-\omegad)\cos\omegad \right)\,,
$$
$$
\frac{\partial \eta}{\partial \omegad}= -\left[ e\cos\omegad - (1
-\sqrt{1-e^2}) \cos(g-2\omegad)\right]\,.
$$

{\it 5. Derivatives of elements of the asteroid orbit with
respect to $x,y,L$.} The formulas relating $x,y,L$, and Keplerian
elements $\omegad, a, e,$ have the following form:
$$
\omegad = \left\{
\begin{array}{cc}
-\arctg\left(\frac{\displaystyle y}{\displaystyle x}\right), & x>0 \\
\rule{0pt}{8mm}\pi-\arctg\left(\frac{\displaystyle y}{\displaystyle x}\right), & x<0 \\
\end{array}
\right. \eqno(A.4)
$$
$$
a=\frac{L^2}{1-\mu}\,,\; e=\frac{\sqrt{s(x^2+y^2)}}{2L}\,,
$$
where
$$
s=4L - x^2 - y^2.
$$
Using formulas (A.4) we find:
$$
\frac{\partial \omegad}{\partial x} = \frac{y}{x^2+y^2}\,,\;
\frac{\partial \omegad}{\partial y} =-\frac{x}{x^2+y^2}\,,\;
\frac{\partial a}{\partial L} =\frac{2L}{1-\mu}\,, \eqno(A.5)
$$
$$
\frac{\partial e}{\partial x} = \frac{x}{2L}
\left(
\sqrt{\frac{s}{x^2+y^2}} - \sqrt{\frac{x^2+y^2}{s}}
\right)\,,
$$
$$
\frac{\partial e}{\partial y} = \frac{y}{2L}
\left(
\sqrt{\frac{s}{x^2+y^2}} - \sqrt{\frac{x^2+y^2}{s}}
\right)\,,
$$
$$
\frac{\partial e}{\partial L} = \frac{\sqrt{x^2+y^2}}{L} \left(
\frac{1}{\sqrt{s}} - \frac{\sqrt{s}}{2L} \right)\,.
$$

{\it 6. Derivatives of $\xi',\eta'$ with respect to  $x,y,L,D.$}
These derivatives are calculated according to the formulas
$$
\frac{\partial \xi'}{\partial x}=\frac{\partial \xi'}{\partial
g'}\frac{\partial g'}{\partial x}\ldots,\;
\frac{\partial\eta'}{\partial x}=\frac{\partial \eta'}{\partial
g'}\frac{\partial g'}{\partial x}\ldots,
$$
where $g'$ is the eccentric longitude of Jupiter.

Differentiating the expressions for coordinates of Jupiter
$$
\xi'=\cos g' - e', \eta'=\sqrt{1-e'^2}\sin g',
$$
we get:
$$
\frac{\partial \xi'}{\partial g'}=-\sin g'\,,\; \frac{\partial
\eta'}{\partial g'}=\sqrt{1-e^2}\cos g'\,.
$$

We find the value of $g'$ from the Kepler's equation
$$
g'-e'\sin g'= \lambda'\,.
$$
Here,
$$
\lambda'=\frac{1}{3}(\lambda - D)=
\frac{1}{3}\left[g-e\sin(g-\omegad) - D\right].
$$
Derivatives of $g'$ with respect to $e,\omegad,D$ have the form:
$$
\frac{\partial g'}{\partial e}=-
\frac{\sin(g-\omegad)}{3(1-e'\cos g')}\,,\; \frac{\partial
g'}{\partial \omegad}= \frac{e\cos(g-\omegad)}{3(1-e'\cos
g')}\,,\;
$$
$$
\frac{\partial g'}{\partial D}=- \frac{1}{3(1-e'\cos g')}\,.
$$
The values of derivatives of $g'$ with respect to we get $x,y,L$
using formulas (A.5):
$$
\frac{\partial g'}{\partial x}=\frac{\partial g'}{\partial
e}\frac{\partial e}{\partial x}+ \frac{\partial g'}{\partial
\omegad}\frac{\partial \omegad}{\partial x}\,,\; \frac{\partial
g'}{\partial y}=\frac{\partial g'}{\partial e}\frac{\partial
e}{\partial y}+ \frac{\partial g'}{\partial
\omegad}\frac{\partial \omegad}{\partial y}\,,\; \frac{\partial
g'}{\partial L}=\frac{\partial g'}{\partial e}\frac{\partial
e}{\partial L}\,.
$$

{\it 7. Derivatives of the mean longitude of an asteroid.}
Differentiating relation (A.2) with respect to eccentric
longitude, we find:
$$
\frac{\partial \lambda}{\partial g}=1 - e\cos(g-\omegad)=
1-\frac{\sqrt{s}}{2L}(x\cos g - y\sin g).
$$
Accordingly:
$$
\frac{\partial^2\lambda}{\partial g\partial x}=
\frac{x}{2L\sqrt{s}}(x\cos g - y\sin g)- \frac{\sqrt{s}}{2L}\cos
g\,,
$$
$$
\frac{\partial^2\lambda}{\partial g\partial y}=
\frac{y}{2L\sqrt{s}}(x\cos g - y\sin g)+ \frac{\sqrt{s}}{2L}\cos
g\,,
$$
$$
\frac{\partial^2\lambda}{\partial g\partial L}= -\frac{1}{L}
\left(\frac{1}{\sqrt{s}} - \frac{\sqrt{s}}{2L}\right) (x\cos g -
y\sin g)
$$

\newpage
\begin{center}

\end{center}
\newpage
\pagestyle{empty}

\unitlength=1mm
\begin{picture}(0,0)
\put(-10,-100){\includegraphics[width=6.0cm,height=9.0cm]{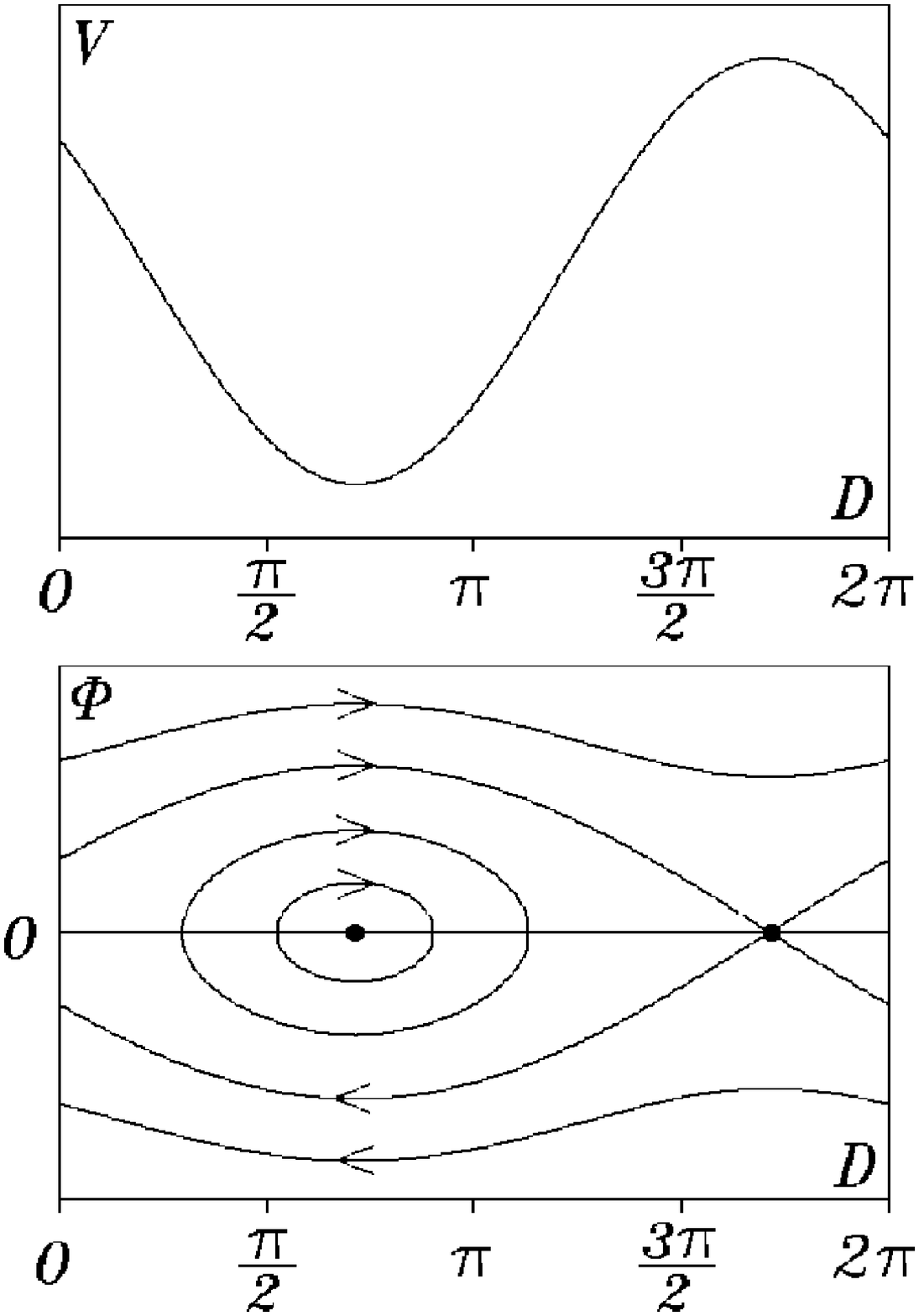}}
\end{picture}

\begin{picture}(0,0)
\put(65,-95){\includegraphics[width=6.0cm,height=9.0cm]{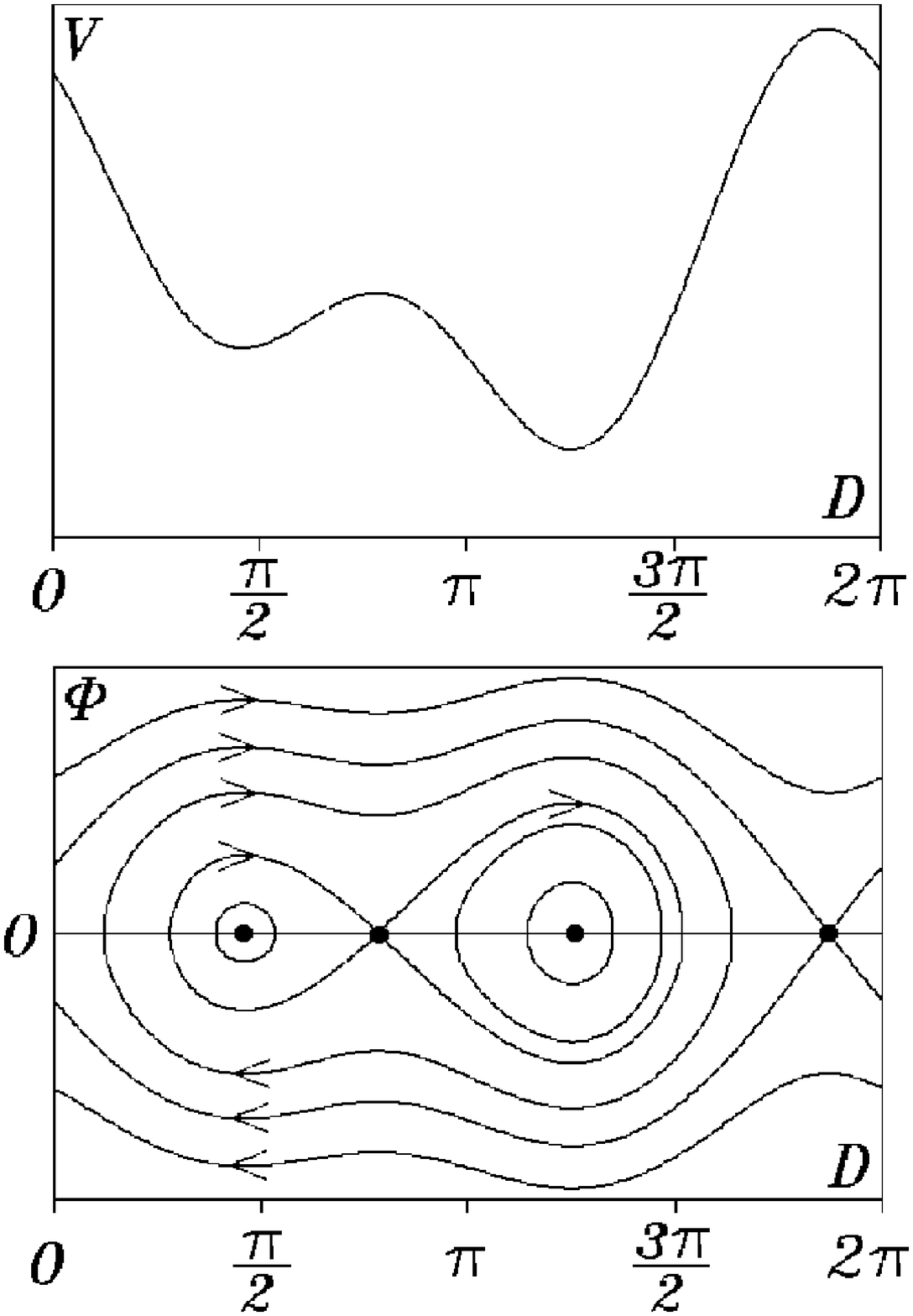}}
\end{picture}

\begin{picture}(0,0)
\put(-12,-190){\includegraphics[width=6.0cm,height=9.0cm]{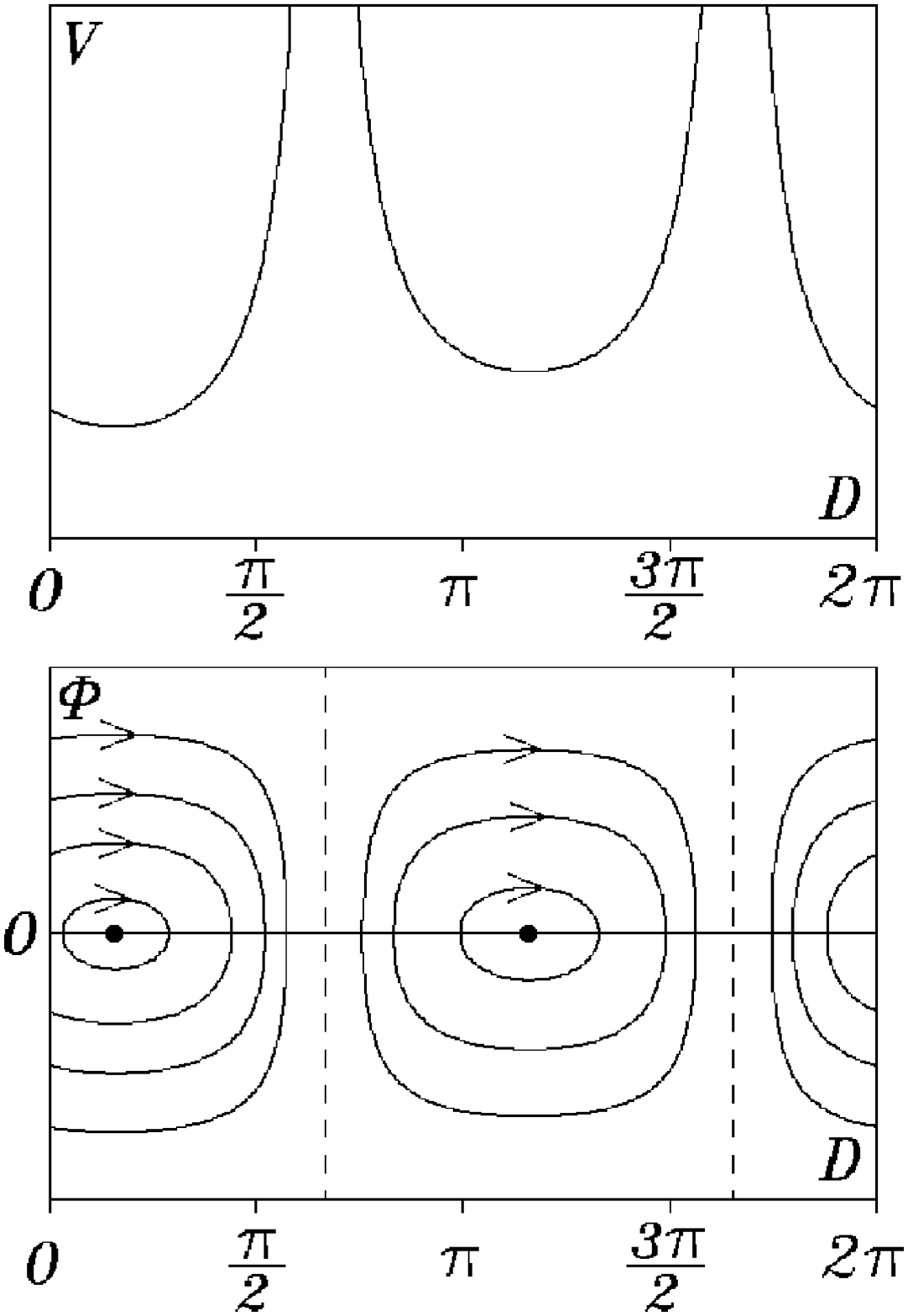}}
\end{picture}

\begin{center}
\vspace*{20cm} Fig. 1. Phase portraits of a fast subsystem:

(a) - $(x,y) \in S_0$, (b) - $(x,y) \in S_1 \cup S_2$, (c) -
$(x,y) \in S_3$.
\end{center}

\newpage
\pagestyle{empty} \unitlength=1mm
\begin{picture}(0,30)
\put(30,-70){\includegraphics[width=7.0cm,height=7.0cm]{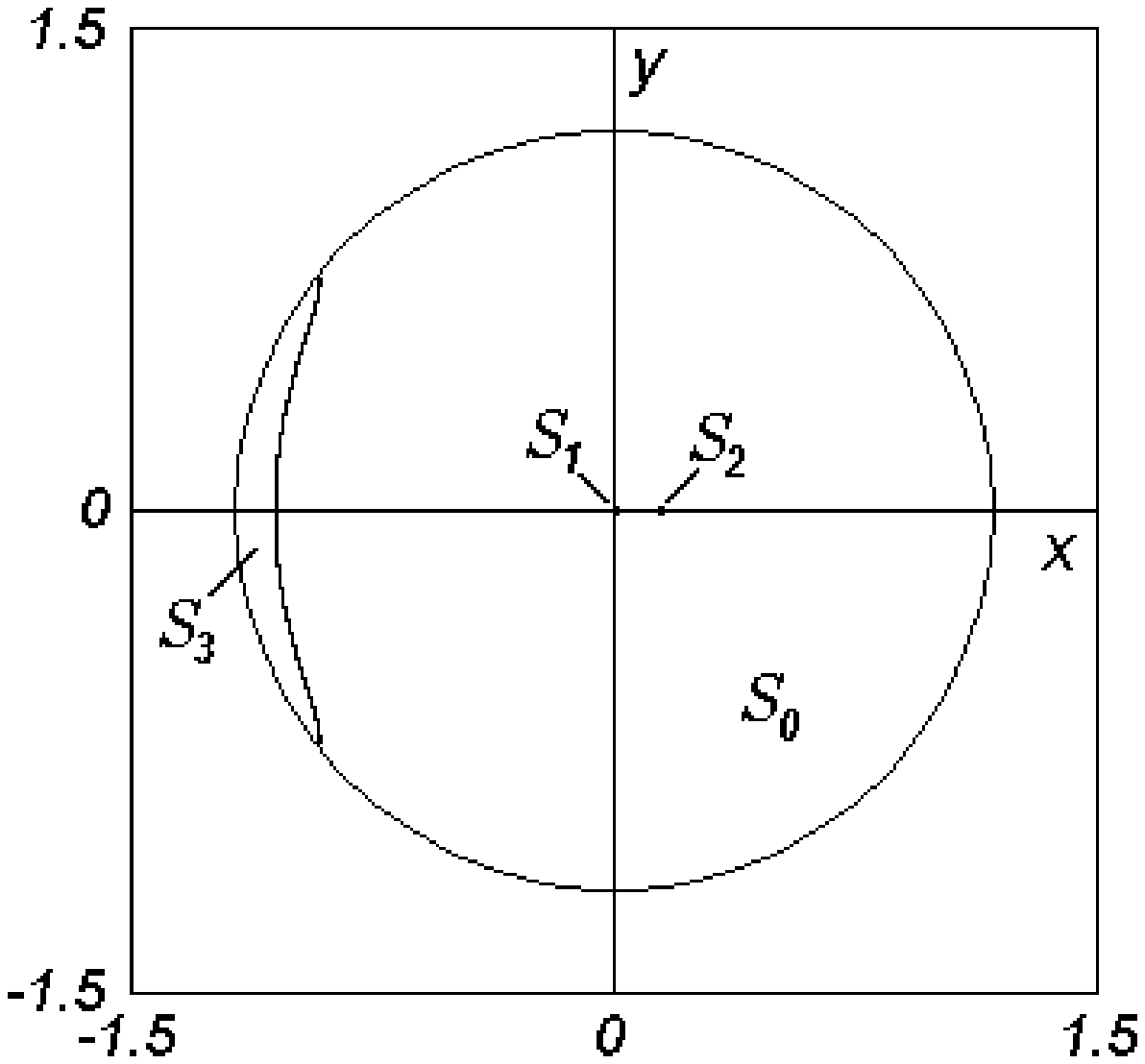}}
\end{picture}

\begin{picture}(0,110)
\put(30,-50){\includegraphics[width=7.0cm,height=7.0cm]{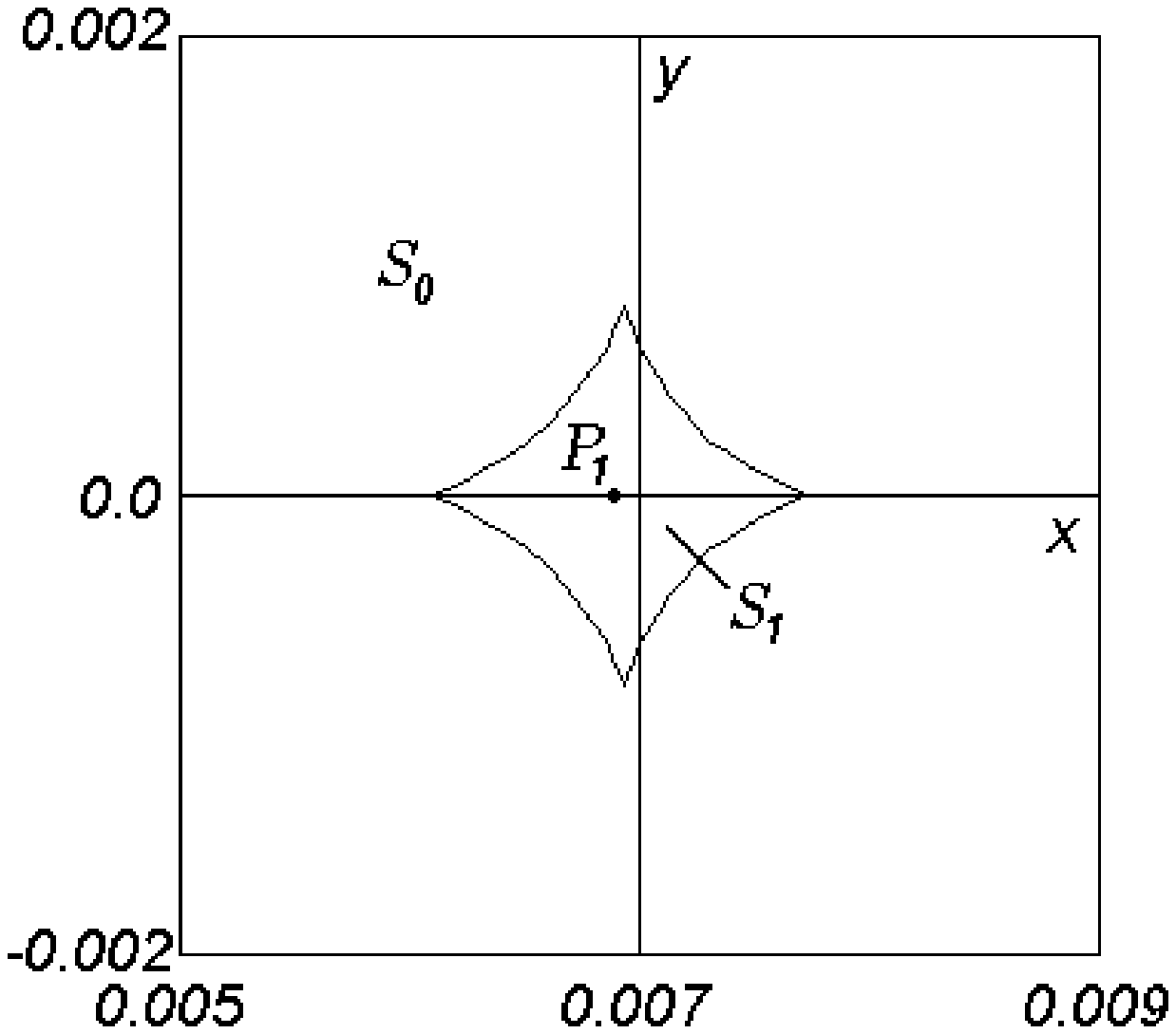}}
\end{picture}

\vspace*{7cm}

Fig. 2. Partitioning of set $S$ into regions of the values of
slow variables $x,y$ corresponding to the dynamics of
qualitatively different types in the "fast" subsystem: (a) -
general view, (b) - enlarged fragment containing region $S_1$.
Region $S_2$ has similar size $(\sim {e'}^3)$ and shape.

\newpage
\pagestyle{empty}

\unitlength=1mm
\begin{picture}(0,50)
\put(30,-30){\includegraphics[width=7.cm,height=7.cm]{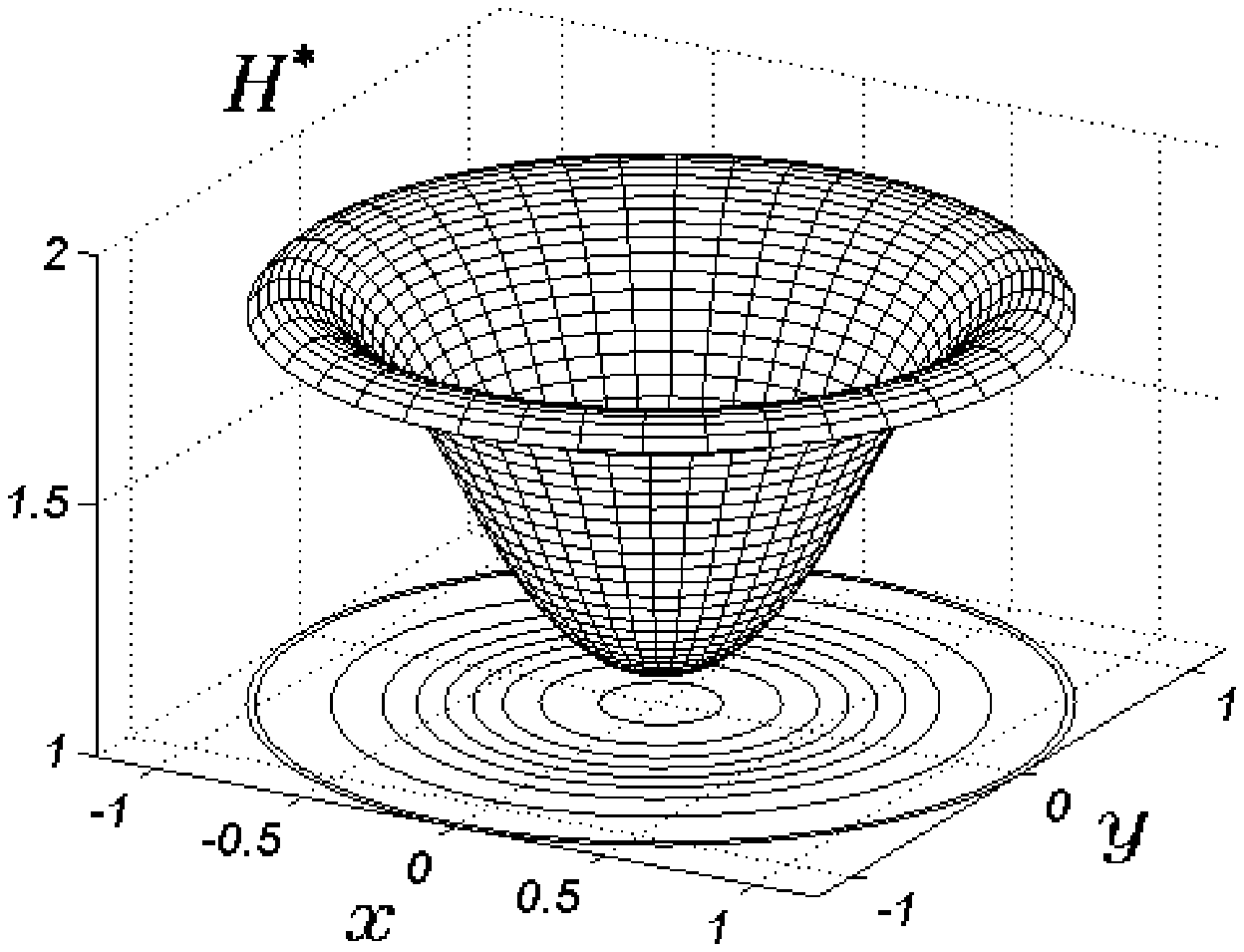}}
\end{picture}

\begin{picture}(0,80)
\put(30,-30){\includegraphics[width=7.cm,height=7.cm]{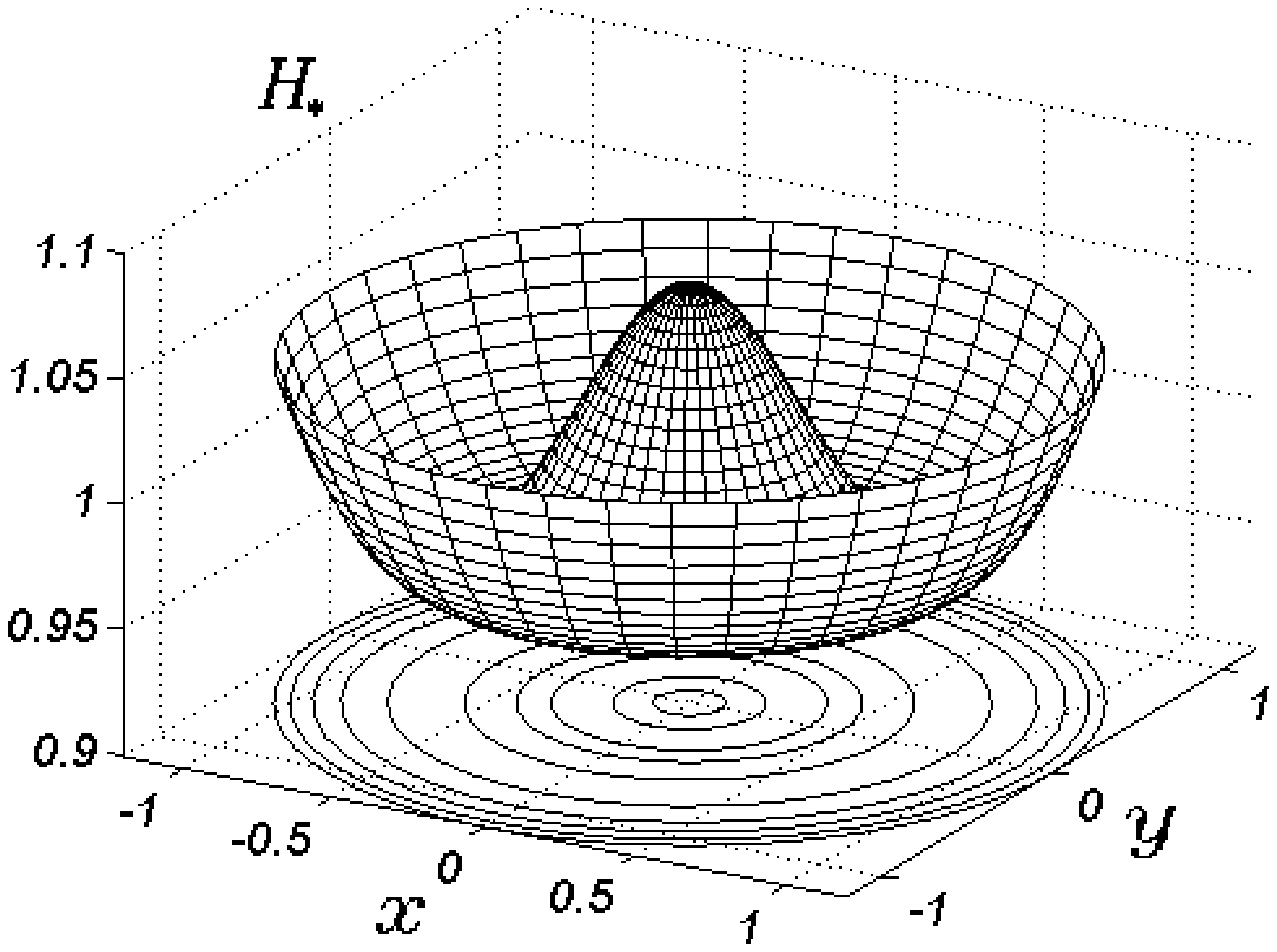}}
\end{picture}

\begin{center}
\vspace*{5cm} Fig. 3. 3D graphs of the functions $H_*(x,y)$ and
$H^*(x,y)$:

(a) $e'=0$
\end{center}
\newpage

\begin{picture}(0,50)
\put(30,-30){\includegraphics[width=7.cm,height=7.cm]{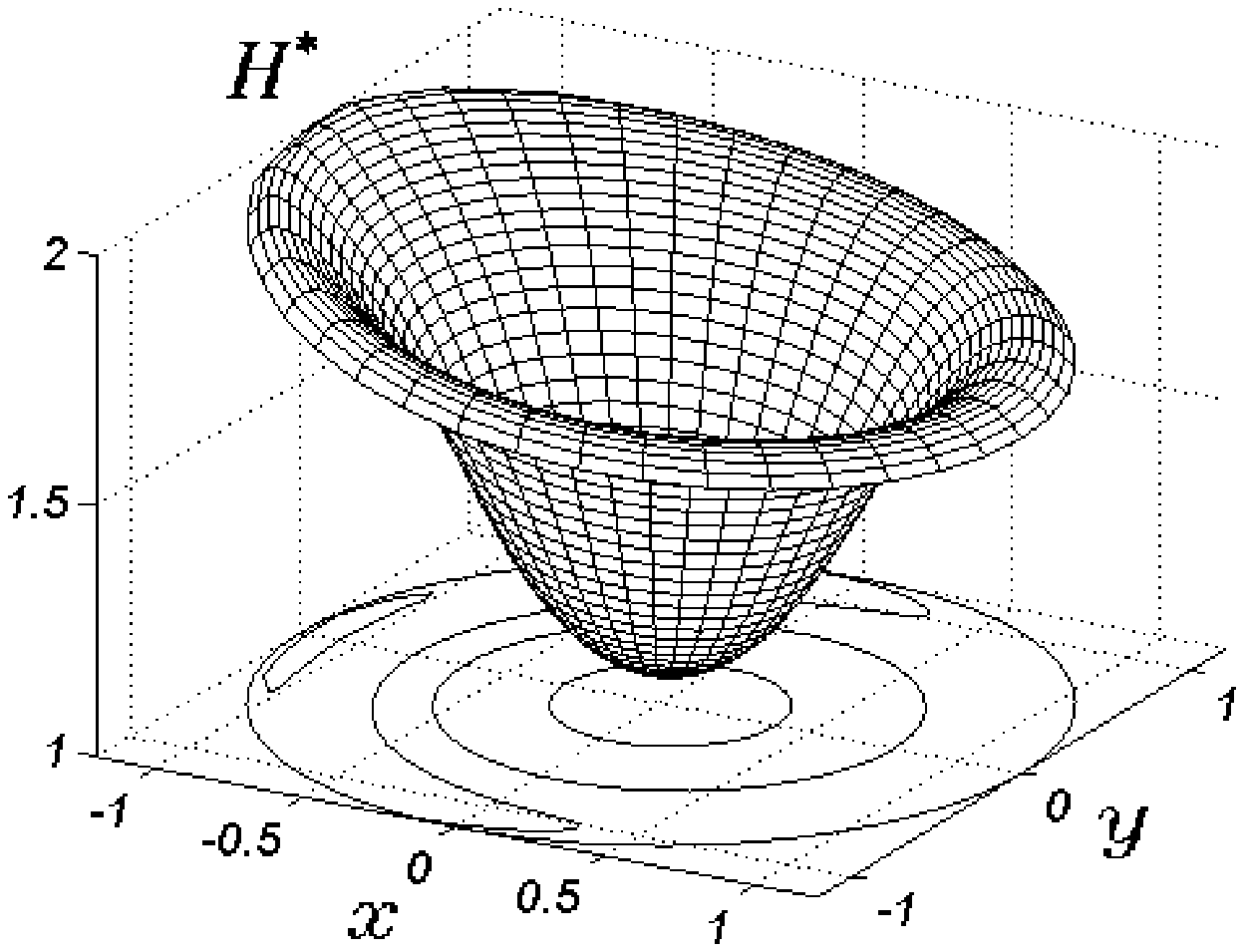}}
\end{picture}

\begin{picture}(0,80)
\put(30,-30){\includegraphics[width=7.0cm,height=7.cm]{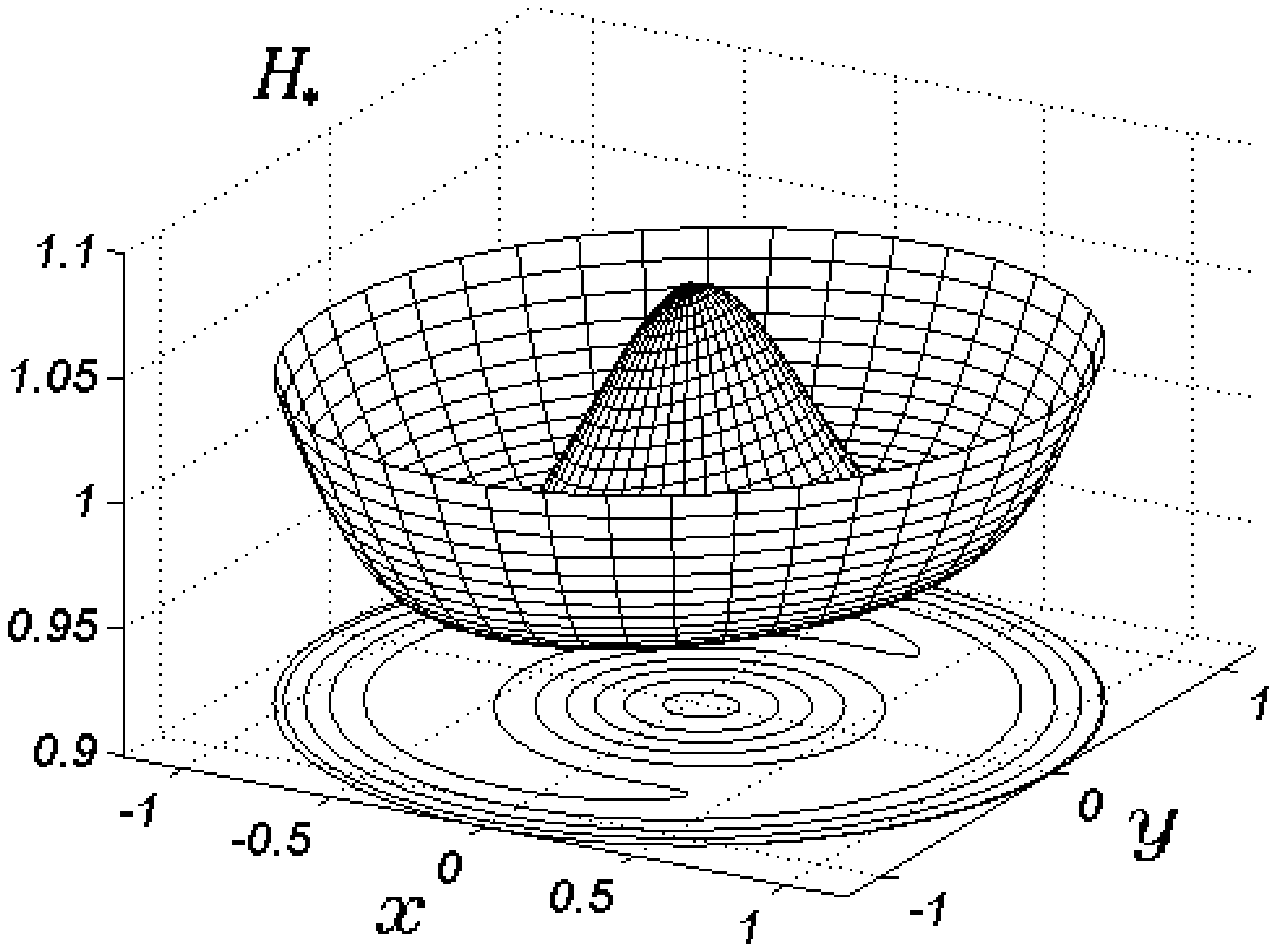}}
\end{picture}

\begin{center}
\vspace*{5cm} Fig. 3. 3D graphs of the functions $H_*(x,y)$ and
$H^*(x,y)$:

(b) $e'=0.02$
\end{center}

\newpage

\begin{picture}(0,50)
\put(30,-30){\includegraphics[width=7.0cm,height=7.cm]{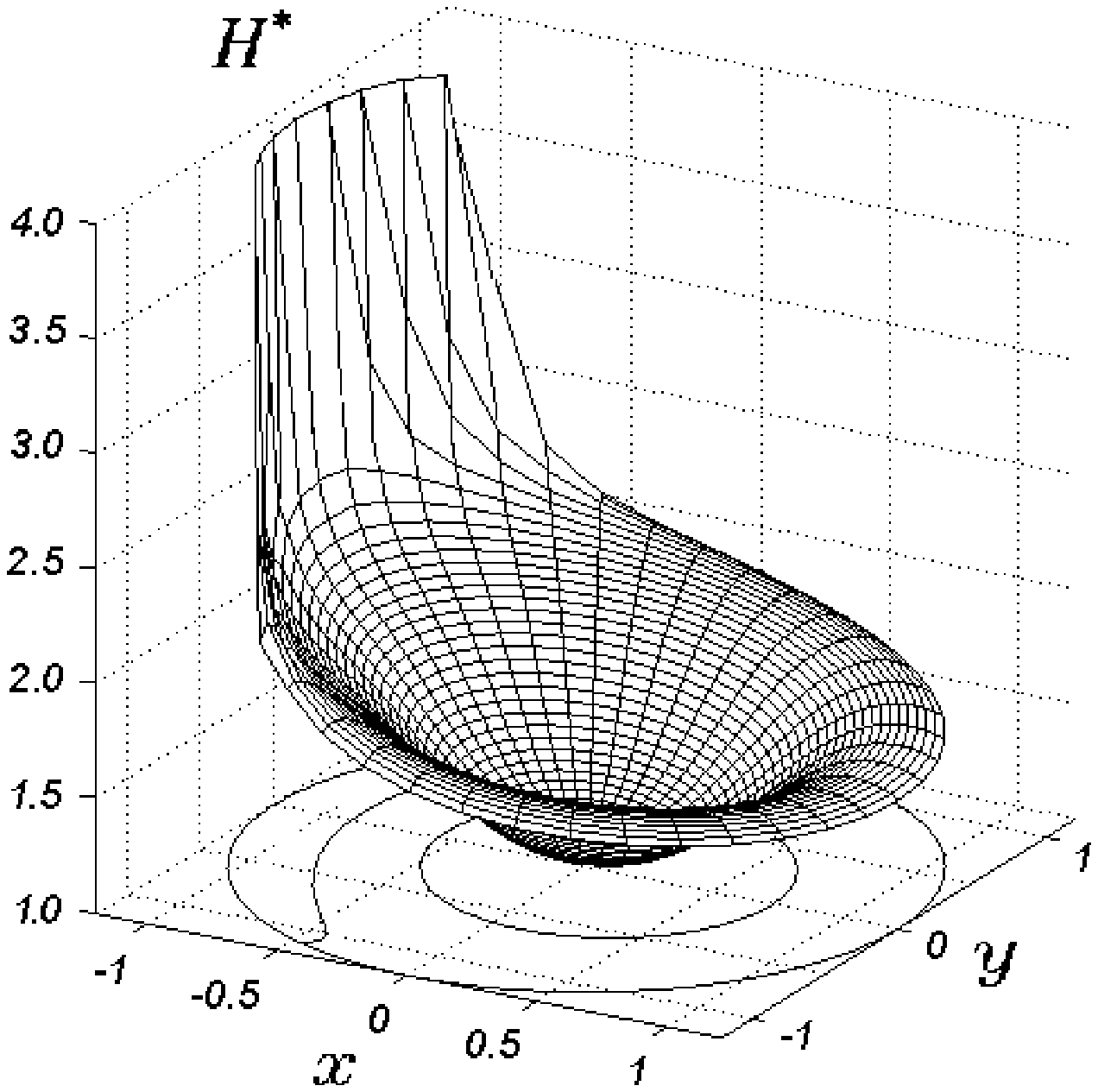}}
\end{picture}

\begin{picture}(0,90)
\put(30,-30){\includegraphics[width=7.0cm,height=7.cm]{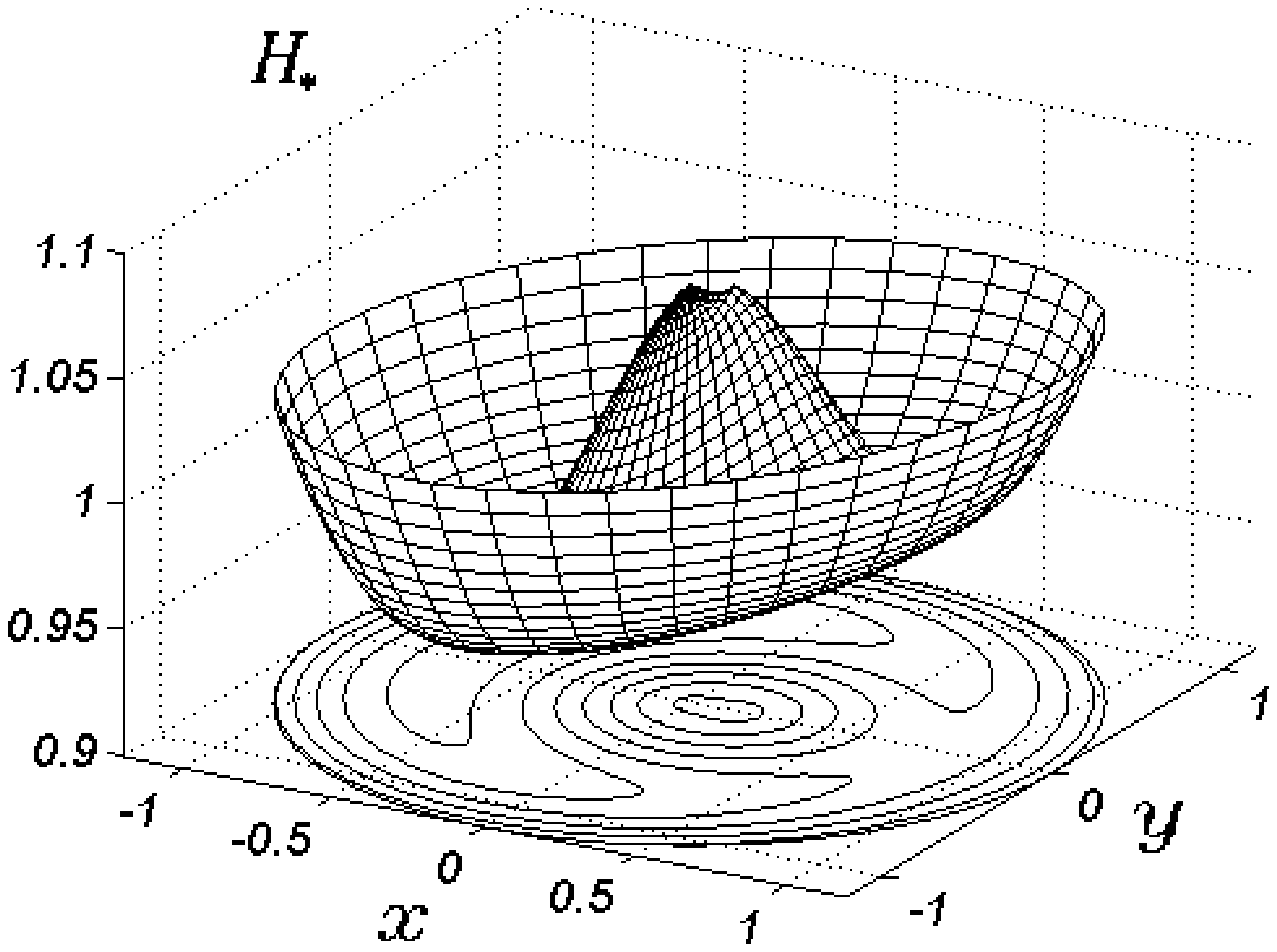}}
\end{picture}

\begin{center}
\vspace*{5cm} Fig. 3. 3D graphs of the functions $H_*(x,y)$ and
$H^*(x,y)$:

(c) $e'=0.048$
\end{center}
\newpage
\pagestyle{empty}

\unitlength=1mm
\begin{picture}(0,0)
\put(30,-100){\includegraphics[width=7.7cm,height=7.7cm]{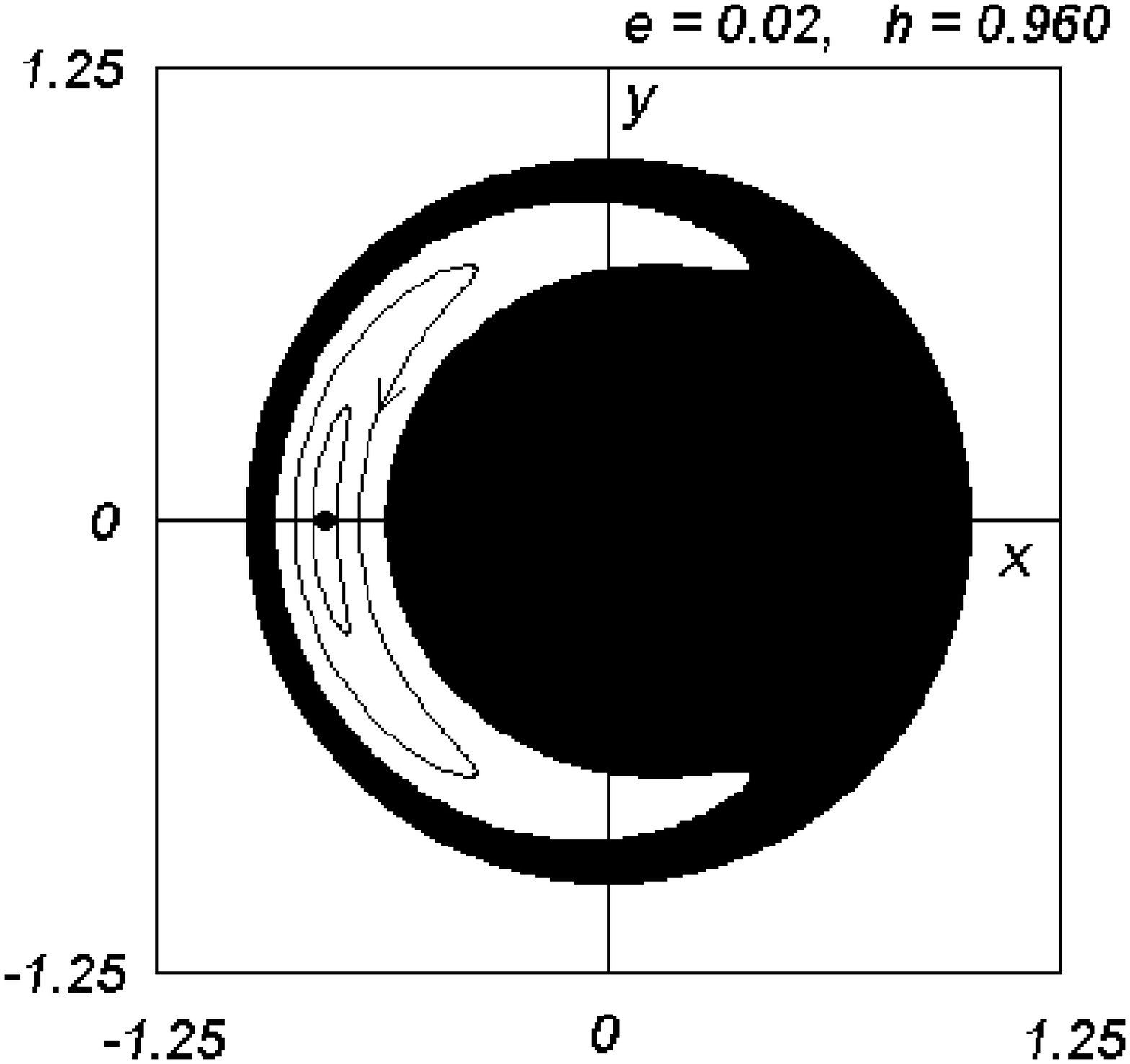}}
\end{picture}

\vspace*{14cm}
\begin{center}
\noindent Fig. 4. Phase portrait of system (6.1) at $h \in
(h_{*min},h_{**}),$ $e'\in (0,e'_*)$.

Here and below, forbidden area $M(h)$ is shown by black color.
\end{center}
\newpage

\begin{picture}(0,0)
\put(30,-100){\includegraphics[width=7.7cm,height=7.7cm]{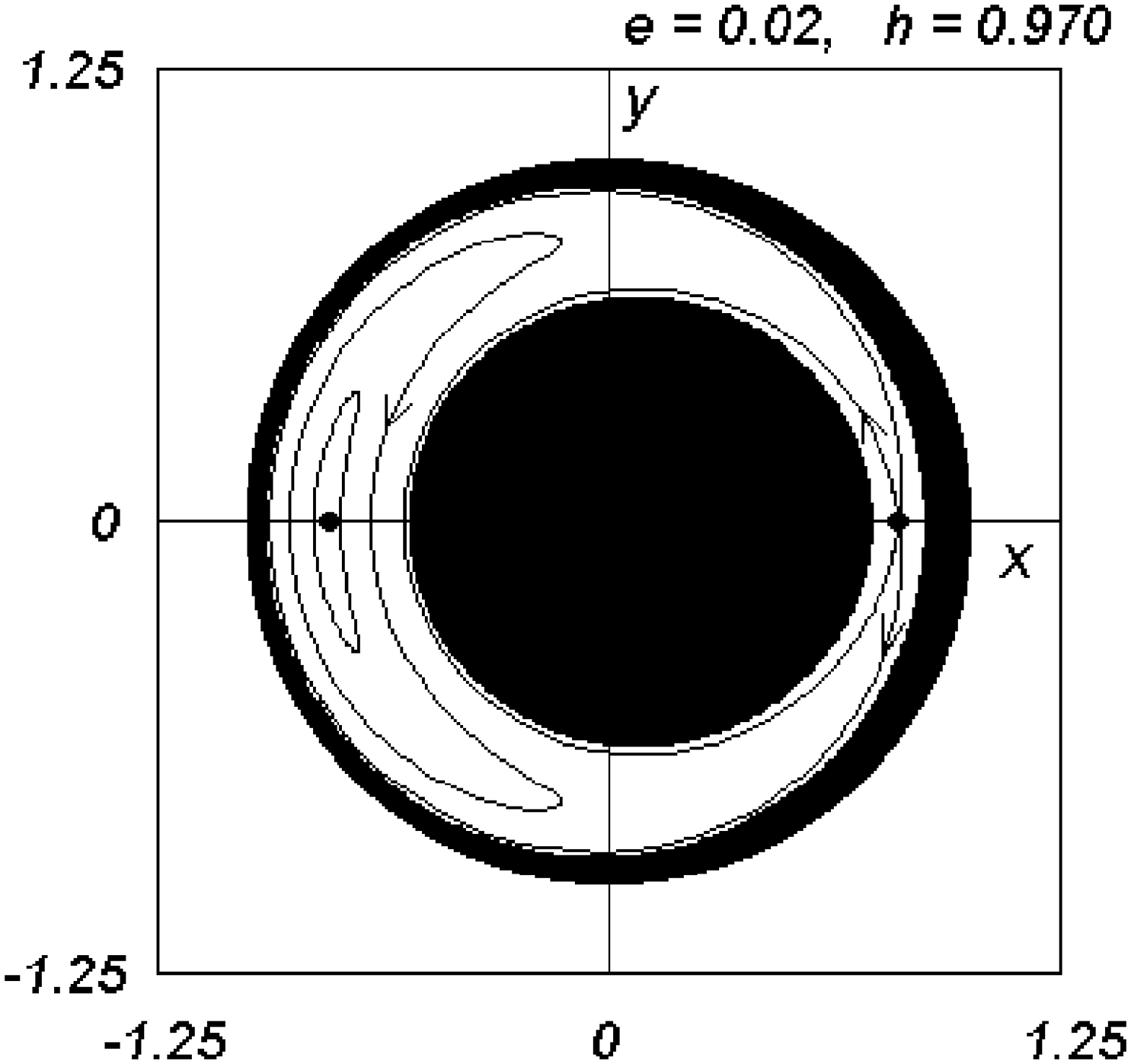}}
\end{picture}

\vspace*{14cm}
\begin{center}
Fig. 5. Phase portrait of system (6.1) at $h \in (h_{**},h_H),$
$e'\in (0,e'_*)$.
\end{center}
\newpage

\begin{picture}(0,0)
\put(30,-75){\includegraphics[width=7.7cm,height=7.7cm]{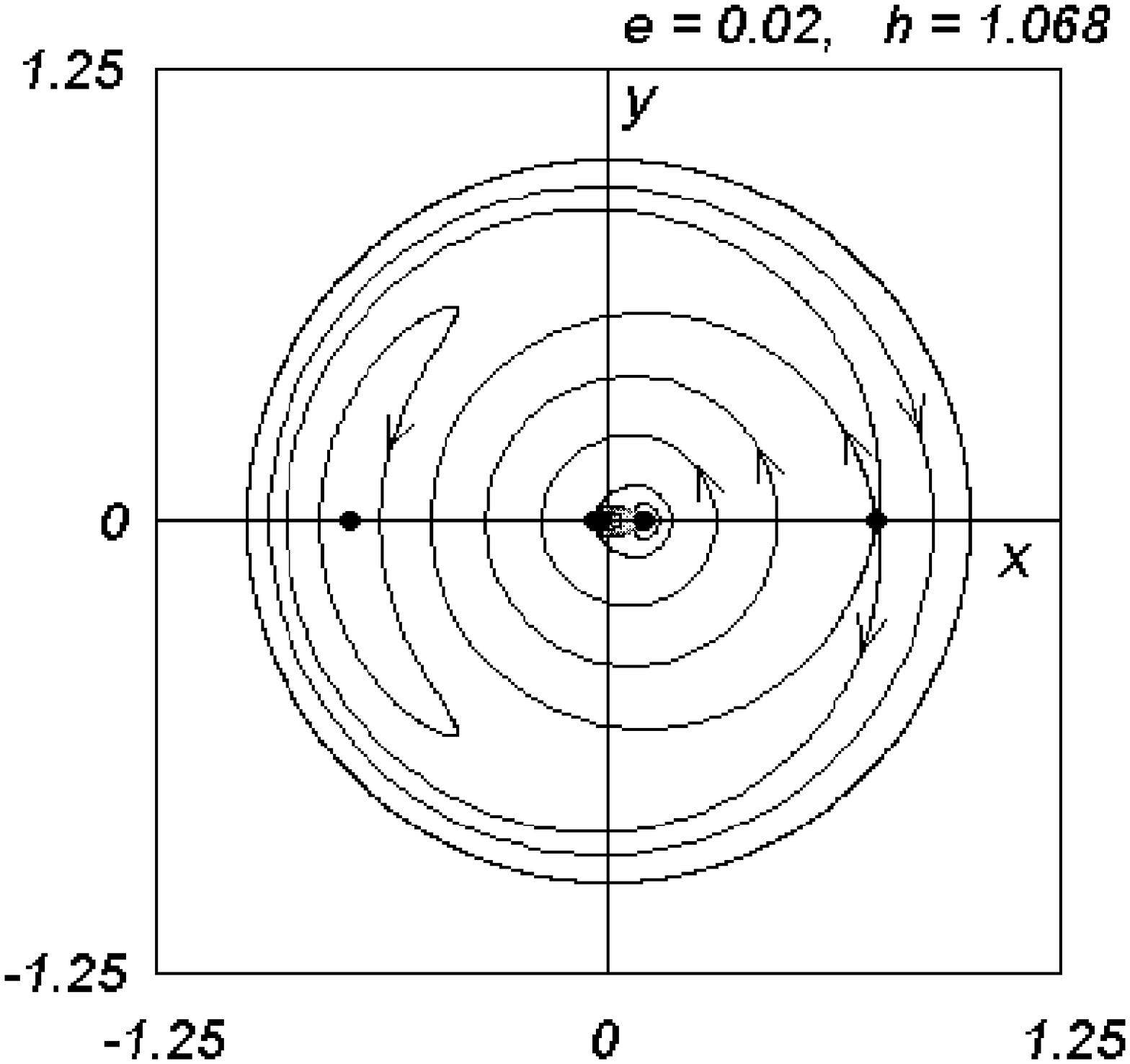}}
\end{picture}

\begin{picture}(0,130)
\put(30,-40){\includegraphics[width=7.7cm,height=7.7cm]{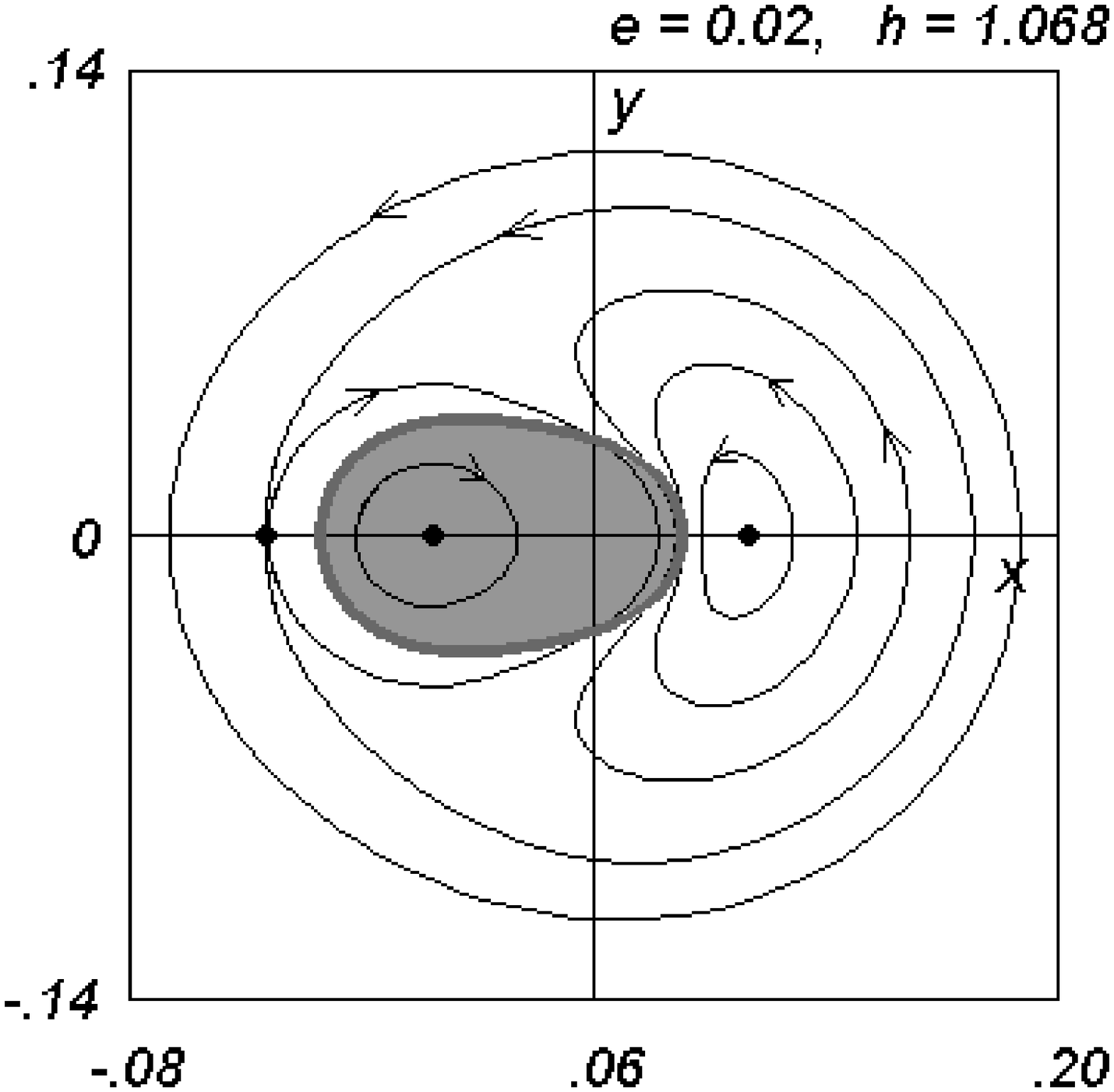}}
\end{picture}
\vspace*{5cm}

Fig. 6. An example of the phase portrait of system (6.1) at $h
\in I_W,$ $e'\in (0,e'_*)$ (a) and an enlarged fragment of this
phase portrait demonstrating the behavior of trajectories in the
Wisdom region $S_W$ (b). Thick line here and below represents
indeterminacy curve $\Gamma(h)$. The region of the values of slow
variables at which the motion in fast subsystem (5.1) has
rotational character at the level ${\cal H}=h$ is marked in gray.

\newpage

\begin{picture}(0,0)
\put(30,-100){\includegraphics[width=7.7cm,height=7.7cm]{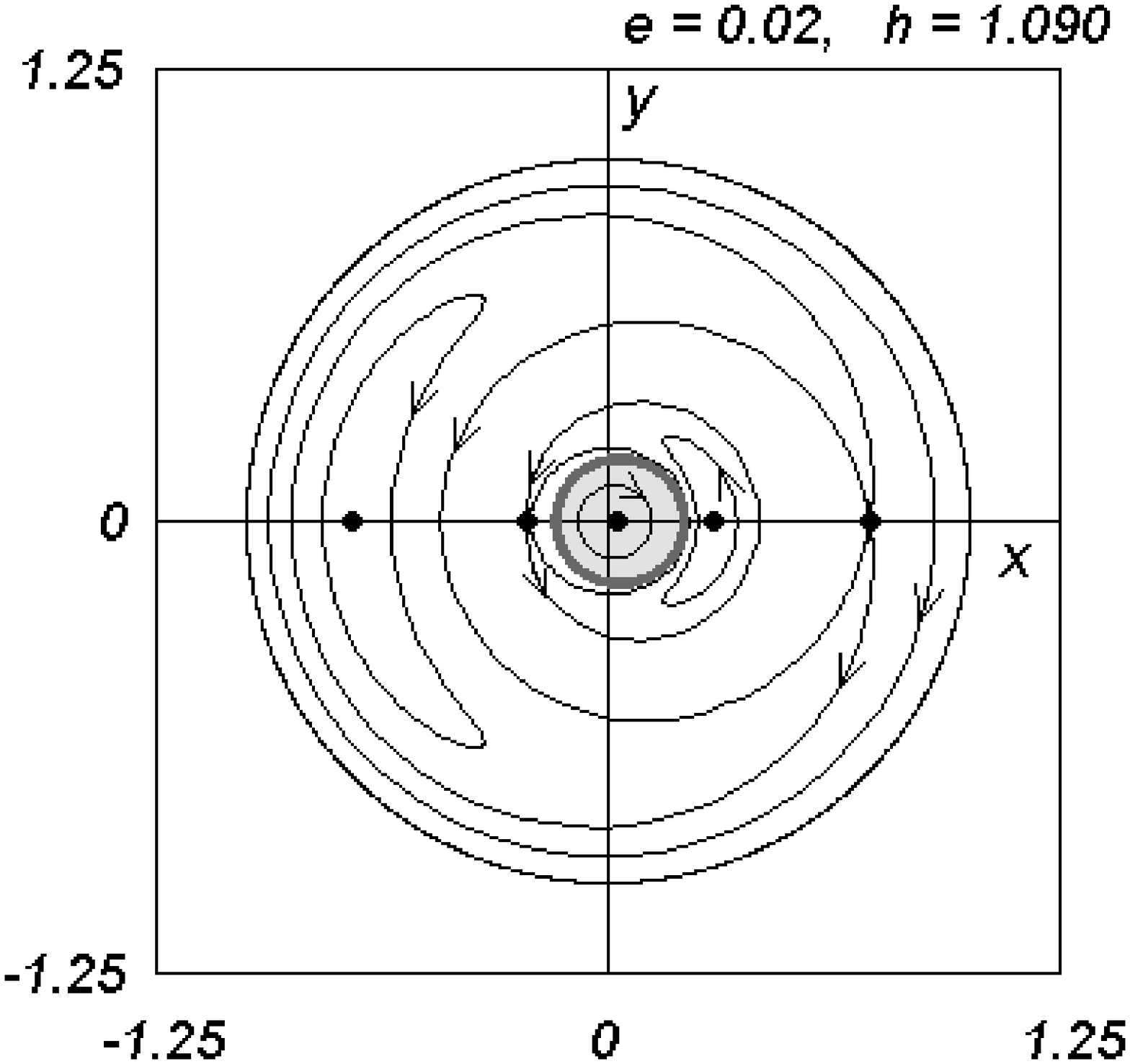}}
\end{picture}

\vspace*{14cm}
\begin{center}
Fig. 7. Phase portrait of system (6.1) after a series of
bifurcations occurring in region $S_W$ when $h$ increases.
\end{center}
\newpage

\begin{picture}(0,0)
\put(30,-100){\includegraphics[width=7.7cm,height=7.7cm]{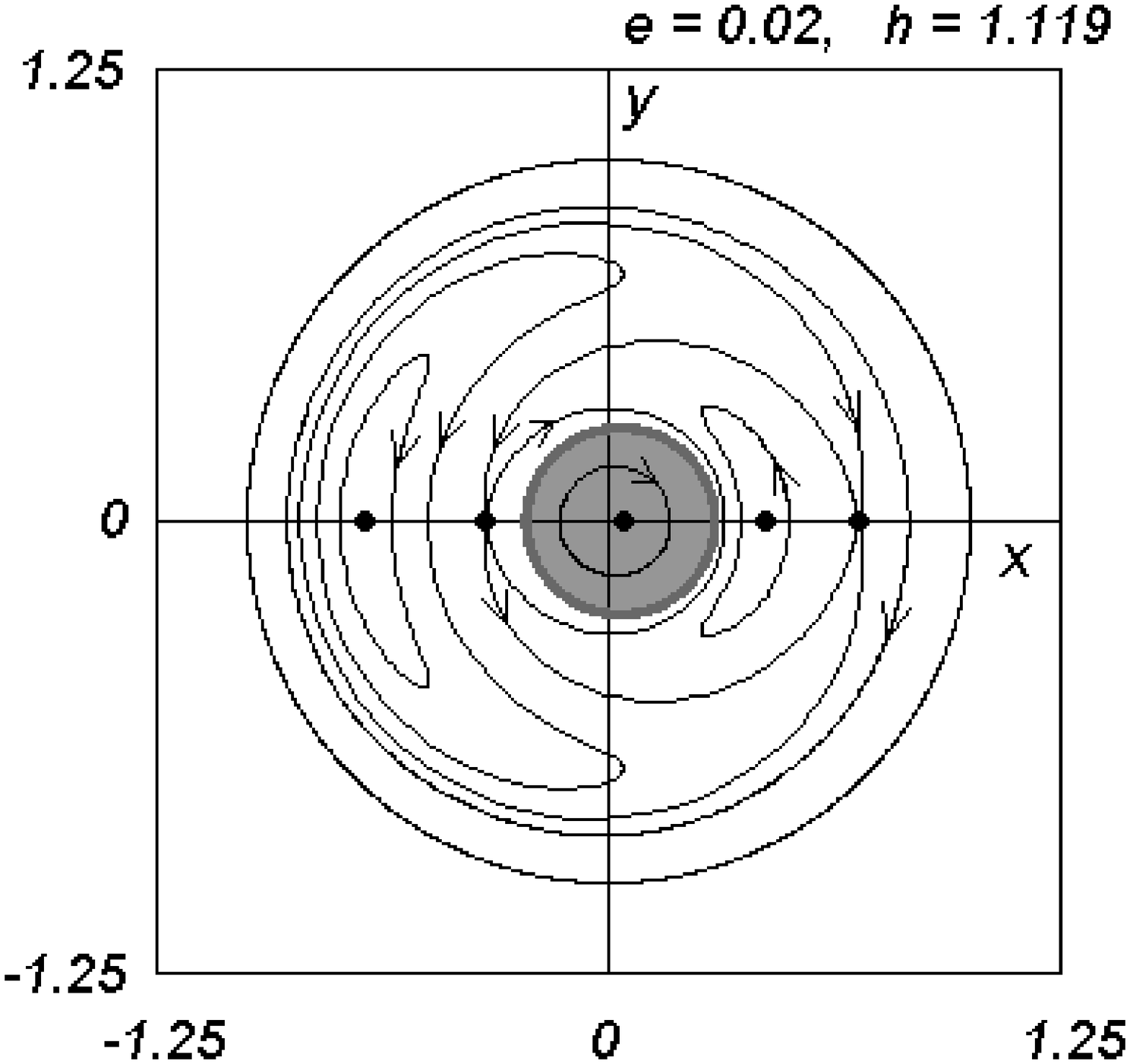}}
\end{picture}

\vspace*{15cm} Fig. 8. Phase portrait of system (6.1) at
bifurcation value of $h$: there are asymptotic solutions with the
different limits at $\tau \to -\infty$ and at $\tau \to +\infty$.
Such solutions disappear at arbitrarily small changes of $h$.

\newpage
\begin{picture}(0,0)
\put(30,-100){\includegraphics[width=7.7cm,height=7.7cm]{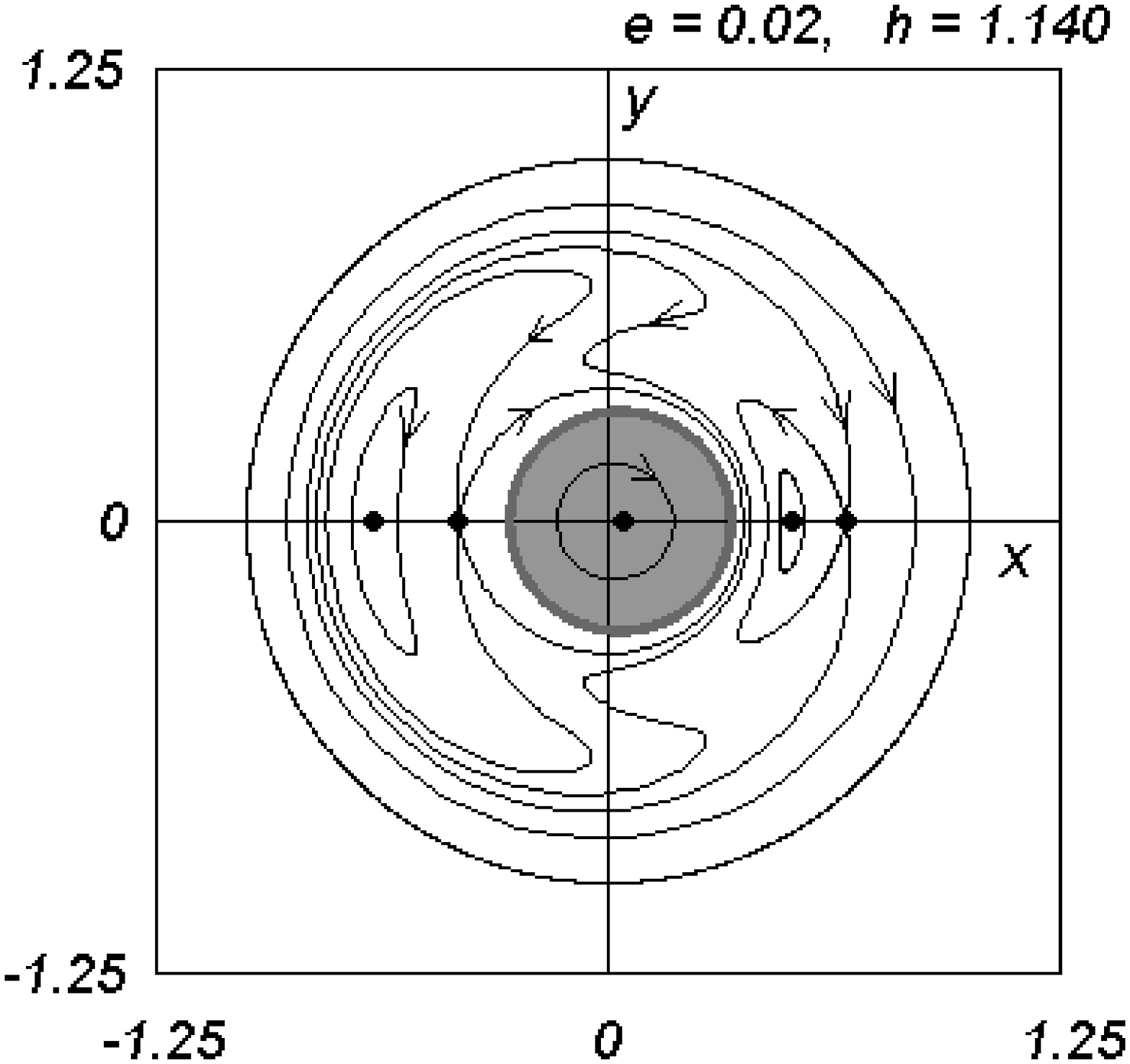}}
\end{picture}

\vspace*{14cm}
\begin{center}
Fig. 9. Phase portrait of system (6.1) after reconnection of
separatrices.
\end{center}
\newpage

\begin{picture}(0,0)
\put(30,-100){\includegraphics[width=7.7cm,height=7.7cm]{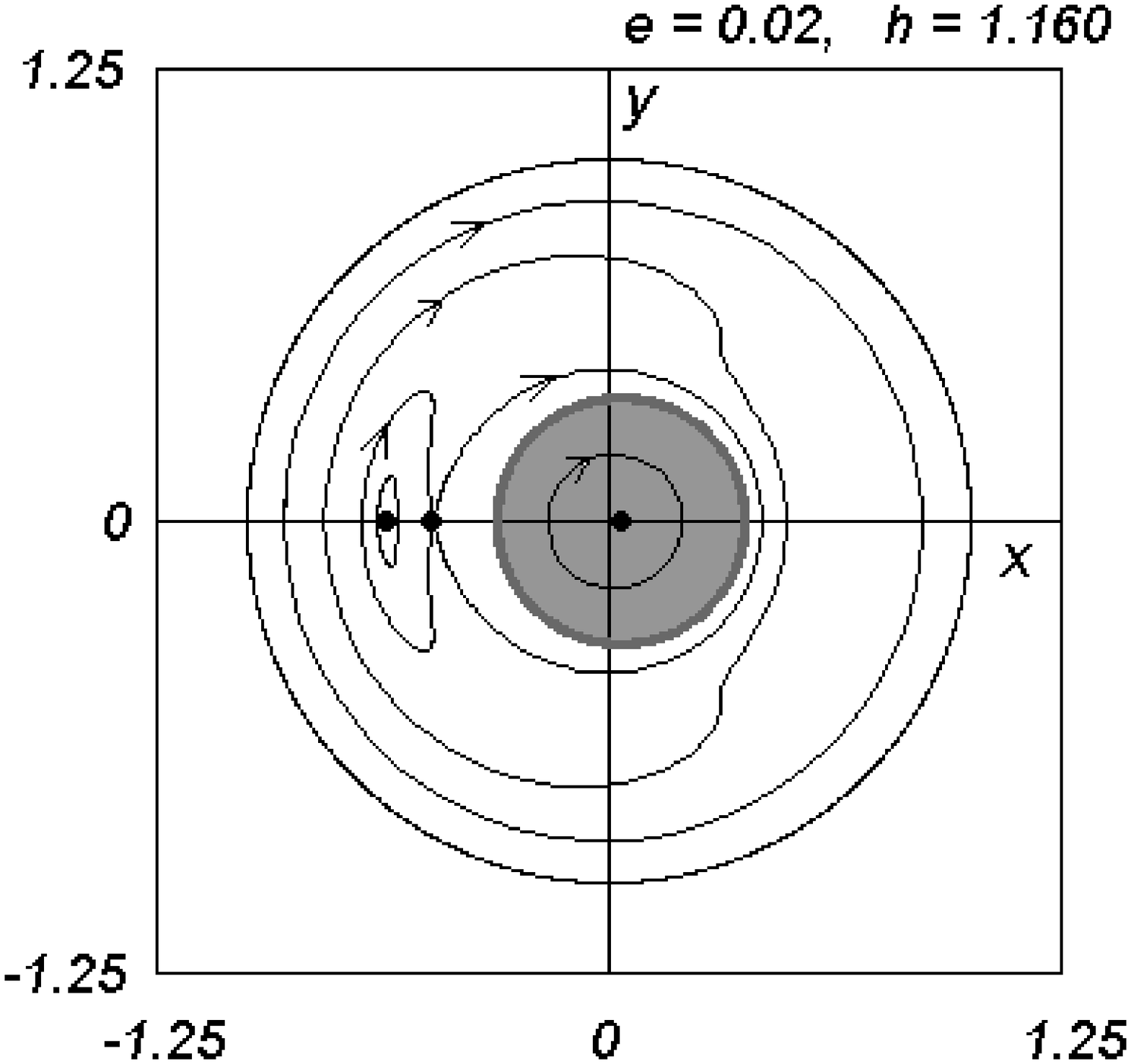}}
\end{picture}

\vspace*{14cm}
\begin{center}
Fig. 10. Phase portrait of system (6.1) after merging of
equilibrium positions in the right-hand half-plane, $e'\in
(0,e'_*)$.
\end{center}

\newpage

\begin{picture}(0,0)
\put(30,-100){\includegraphics[width=7.7cm,height=7.7cm]{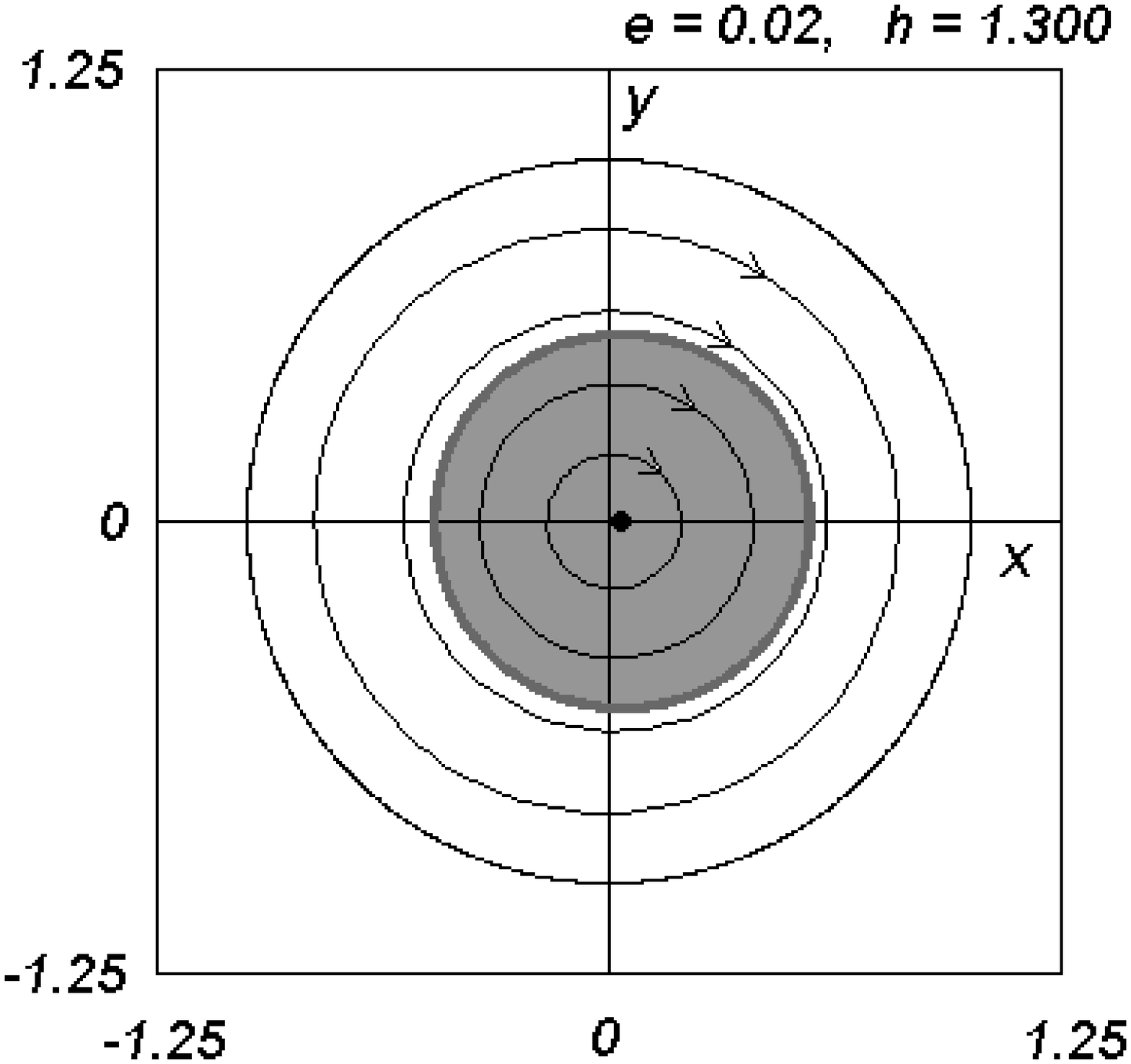}}
\end{picture}

\vspace*{14cm}
\begin{center}
Fig. 11. Phase portrait of system (6.1) after merging of
equilibrium positions in the left-hand half-plane, $e'\in
(0,e'_*)$.
\end{center}
\newpage

\begin{picture}(0,0)
\put(30,-100){\includegraphics[width=7.7cm,height=7.7cm]{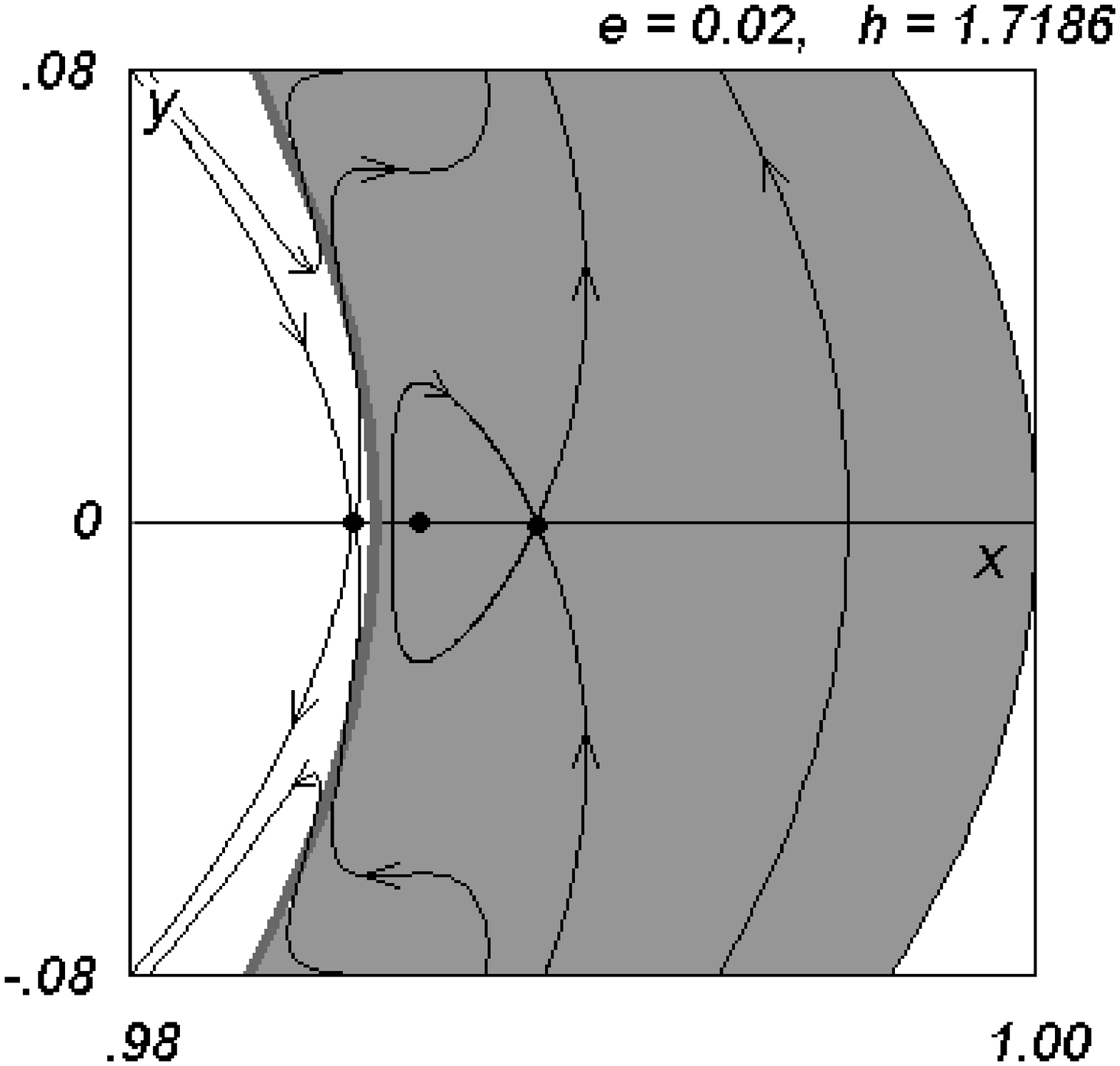}}
\end{picture}

\vspace*{14cm}
\begin{center}
Fig. 12. A fragment of the phase portrait of system (6.1) at $h <
h^{**},$ $|h-h^{**}| \ll 1,$ $e'\in (0,e'_*)$.
\end{center}
\newpage

\begin{picture}(0,20)
\put(30,-40){\includegraphics[width=7.0cm,height=7.0cm]{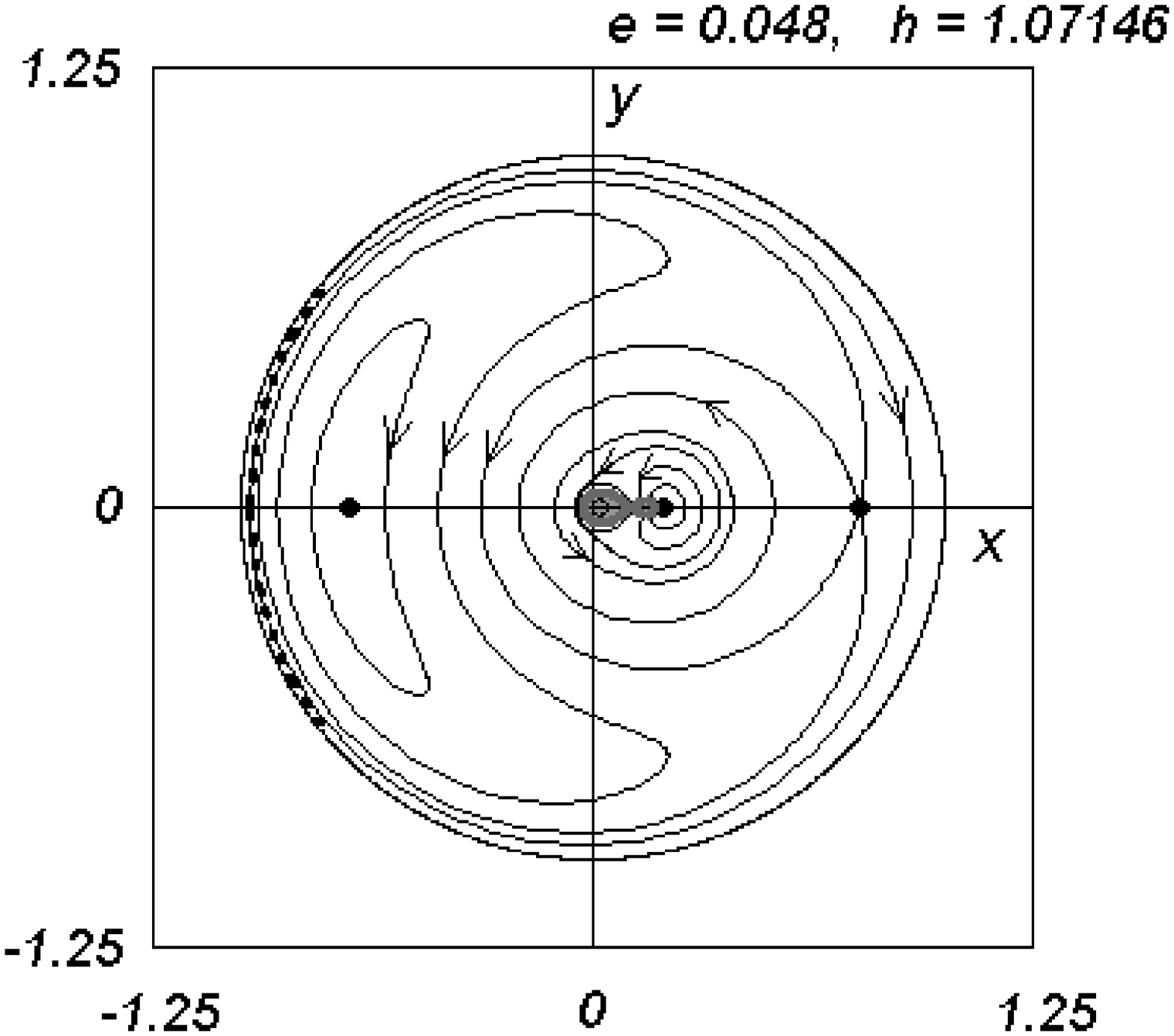}}
\end{picture}

\begin{picture}(0,80)
\put(30,-40){\includegraphics[width=7.0cm,height=7.0cm]{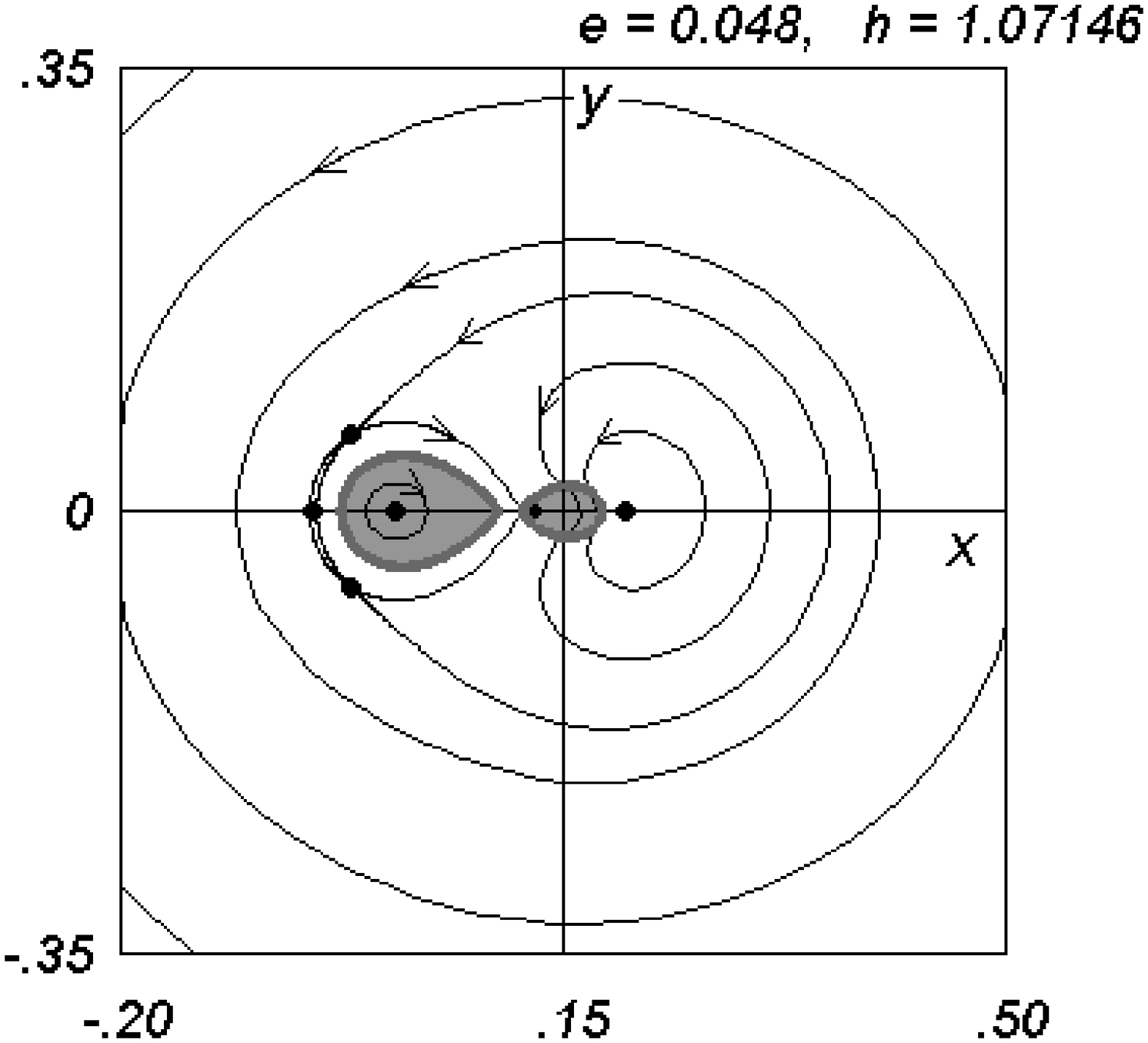}}
\end{picture}

\begin{picture}(0,80)
\put(30,-40){\includegraphics[width=7.0cm,height=7.0cm]{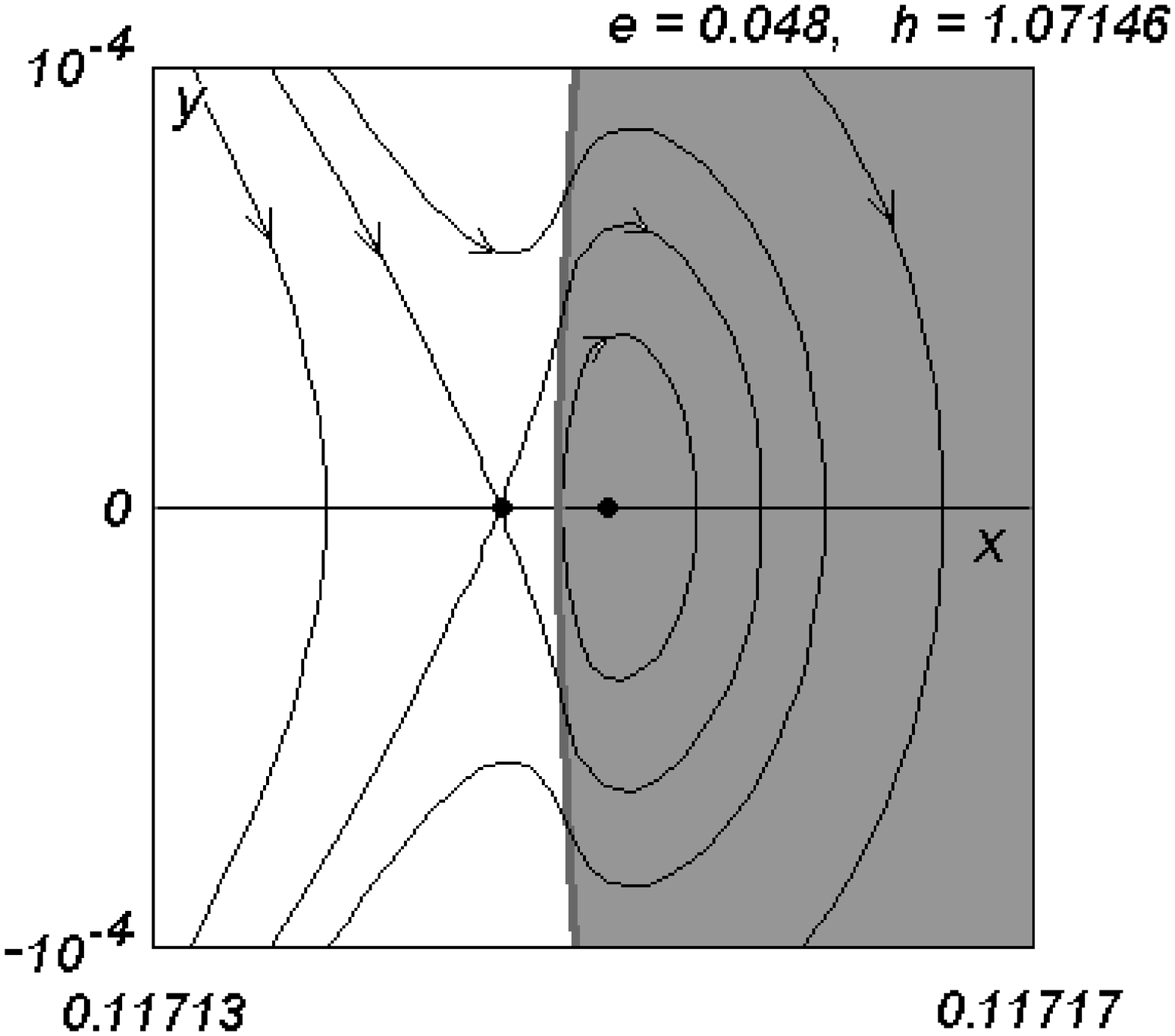}}
\end{picture}

\vspace*{5cm} Fig. 13. Phase portrait of system (6.1) at $h \in
I_W$ $(e'>e'_*)$ (a) and its enlarged fragments (b and c). Dashed
line shows the position of the boundary between regions $S_0$ and
$S_3$.
\newpage

\begin{picture}(0,0)
\put(30,-100){\includegraphics[width=7.7cm,height=7.7cm]{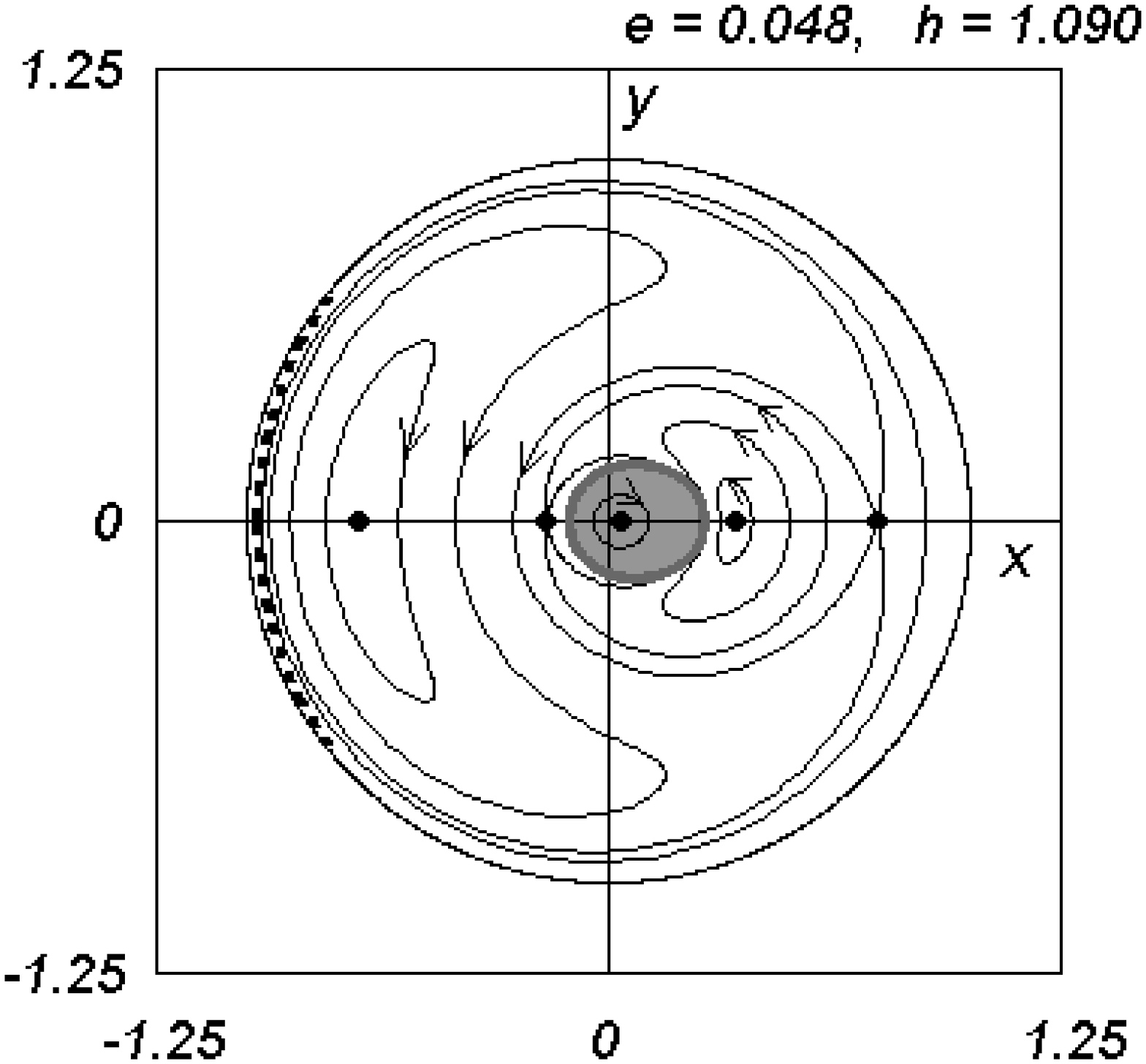}}
\end{picture}

\vspace*{14cm}
\begin{center}
Fig. 14. Phase portrait of system (6.1) before reconnection of
separatrices $(e'>e'_*)$.
\end{center}
\newpage

\begin{picture}(0,0)
\put(30,-100){\includegraphics[width=7.7cm,height=7.7cm]{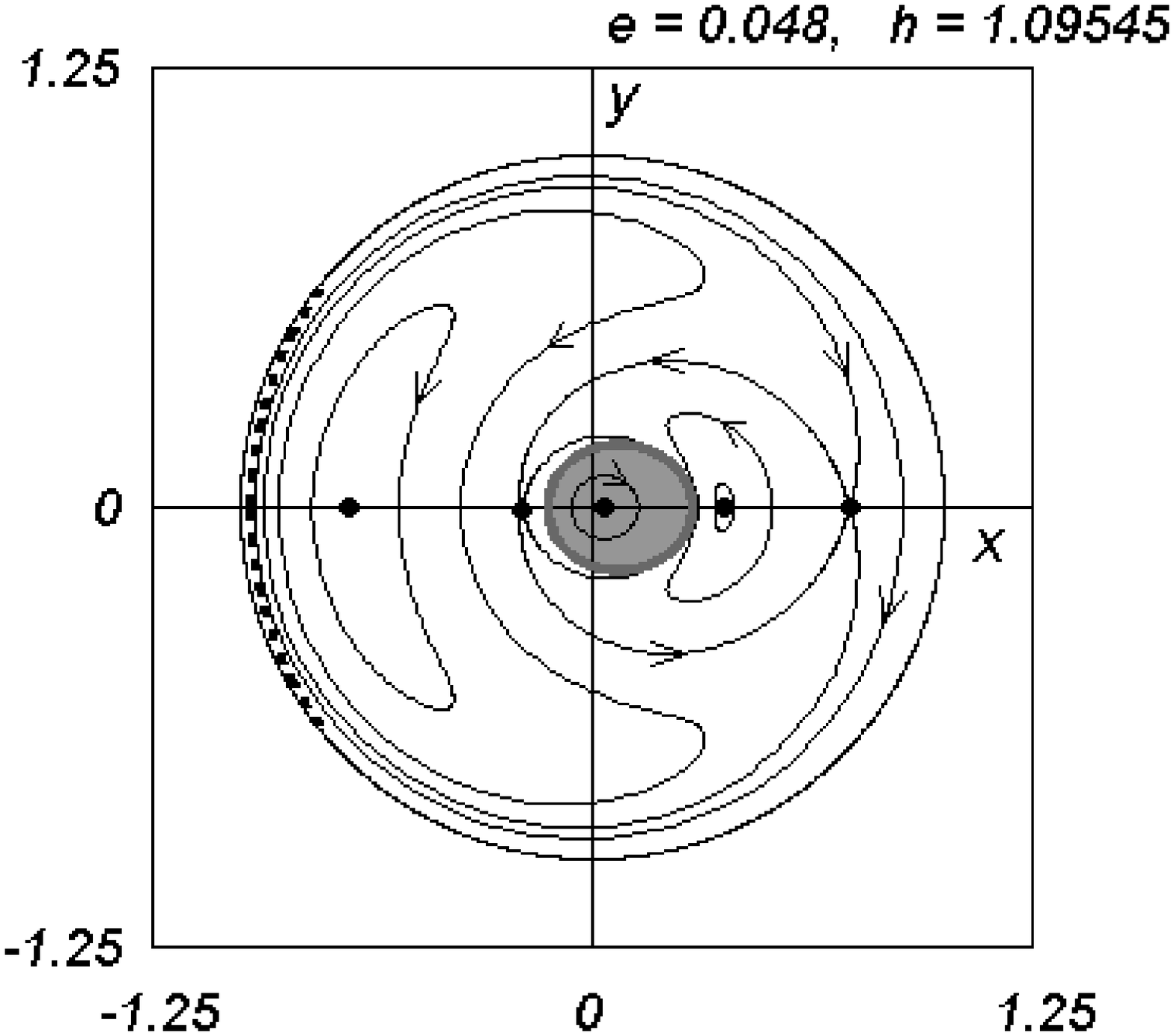}}
\end{picture}
\vspace*{14cm}
\begin{center}
Fig. 15. Phase portrait of system (6.1) for bifurcation value of
$h$ $(e'>e'_*)$.
\end{center}
\newpage
\begin{picture}(0,0)
\put(30,-100){\includegraphics[width=7.7cm,height=7.7cm]{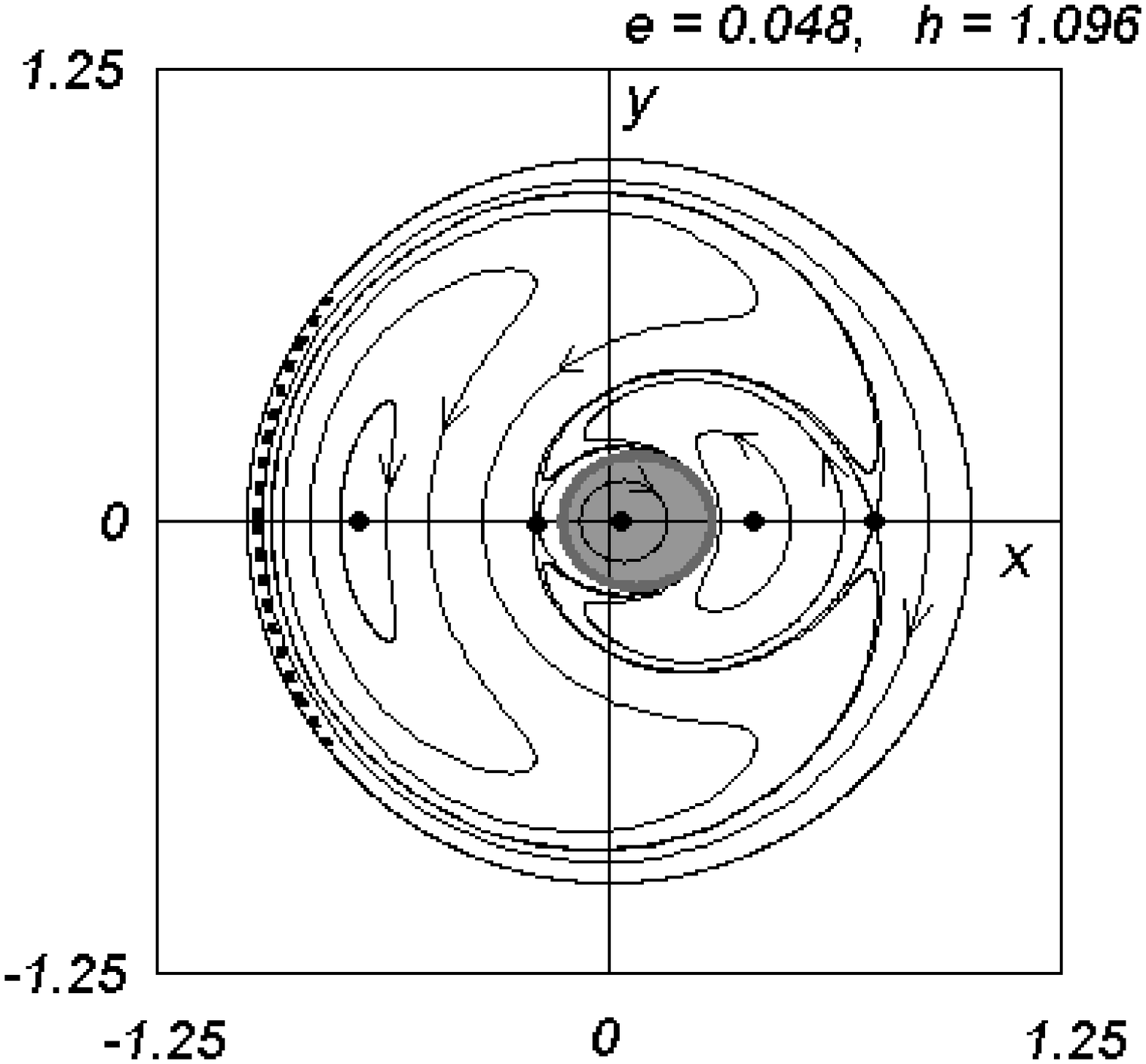}}
\end{picture}
\vspace*{14cm}
\begin{center}
Fig. 16. Phase portrait of system (6.1) after reconnection of
separatrices $(e'>e'_*)$.
\end{center}
\newpage \pagestyle{empty}

\unitlength=1mm
\begin{picture}(0,25)
\put(15,-70){\includegraphics[width=10.0cm,height=8.0cm]{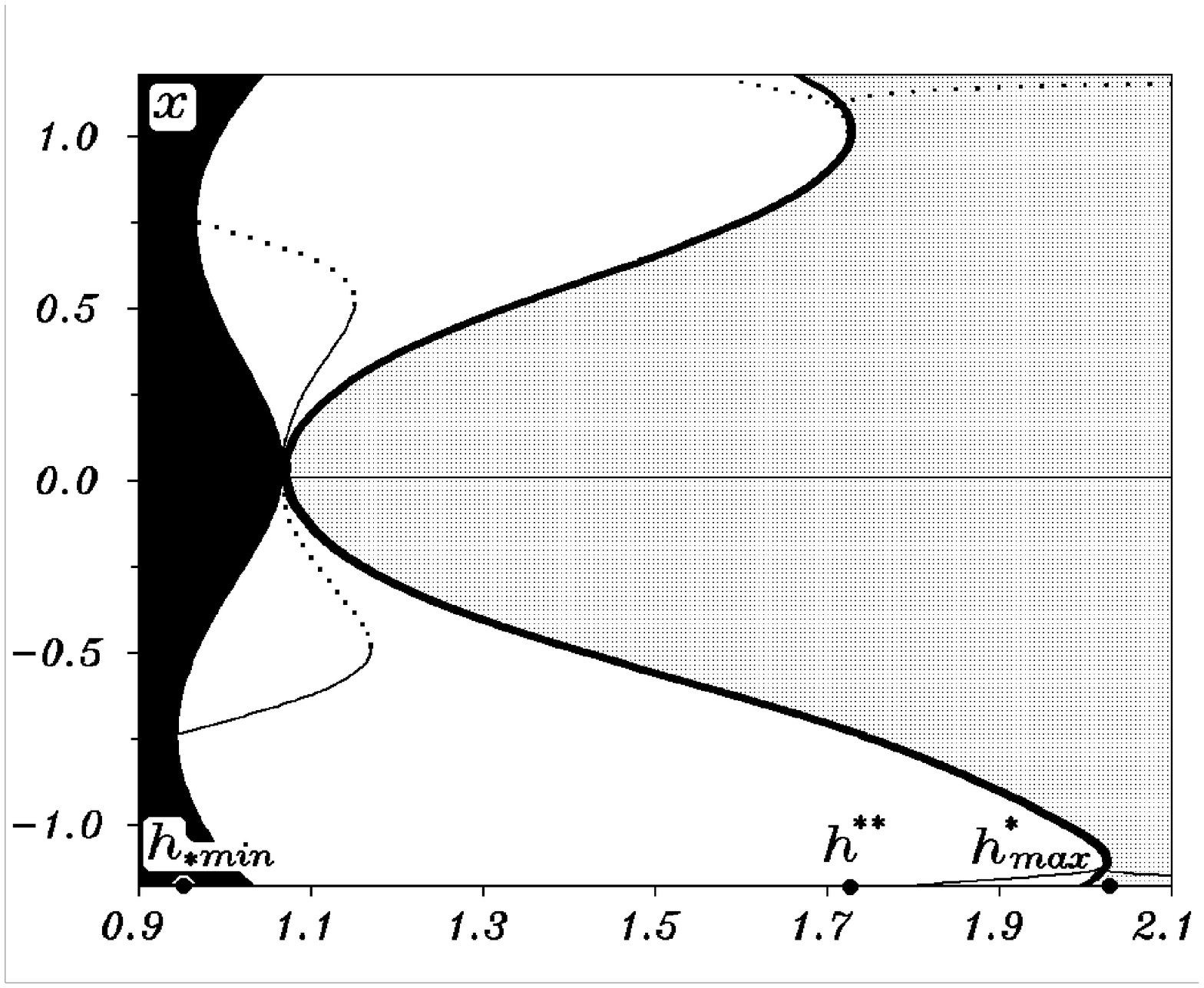}}
\end{picture}

\begin{picture}(0,100)
\put(10,-14){\includegraphics[width=5.0cm,height=4.0cm]{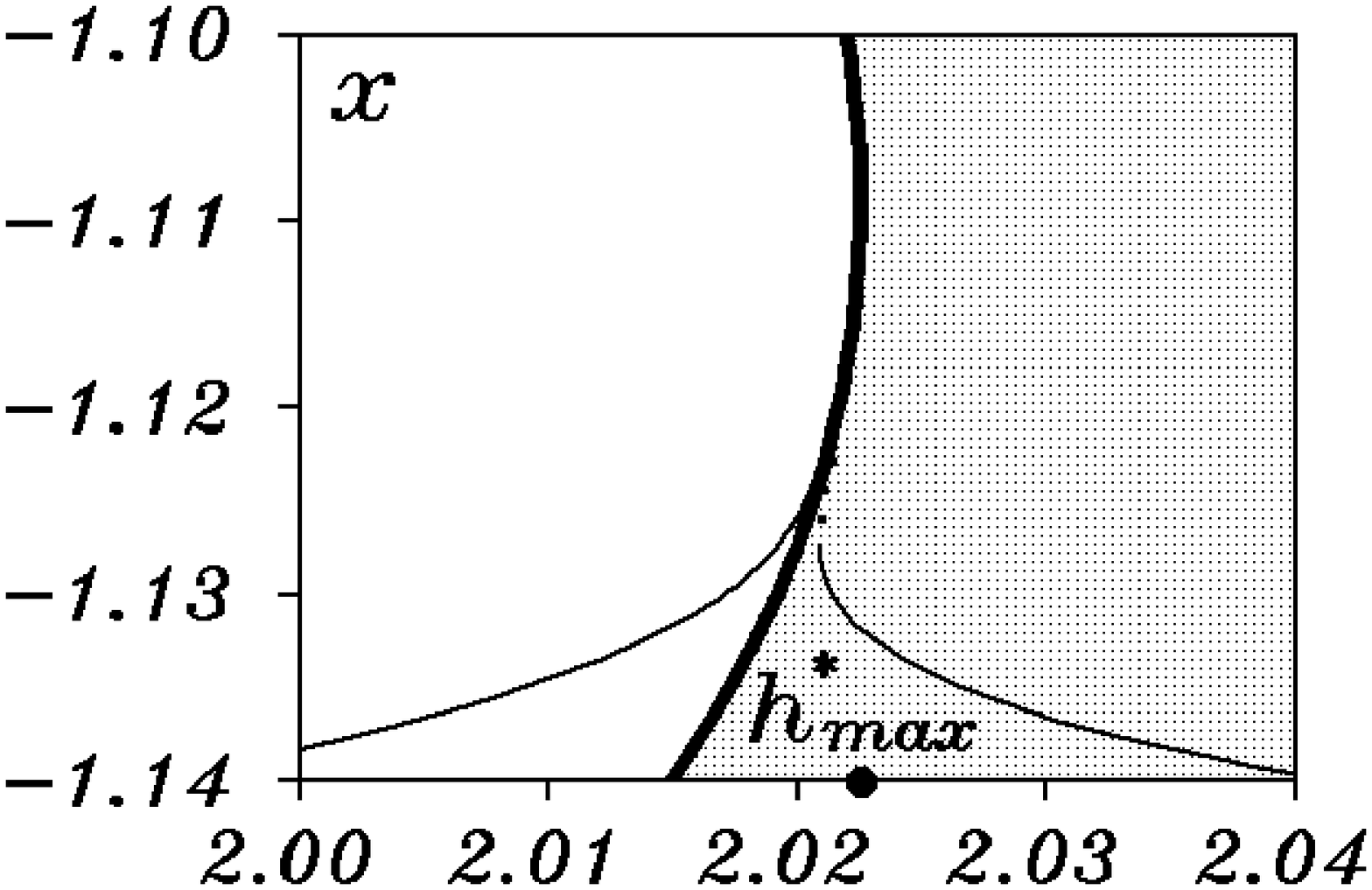}}
\end{picture}

\begin{picture}(0,0)
\put(70,-8){\includegraphics[width=5.0cm,height=4.0cm]{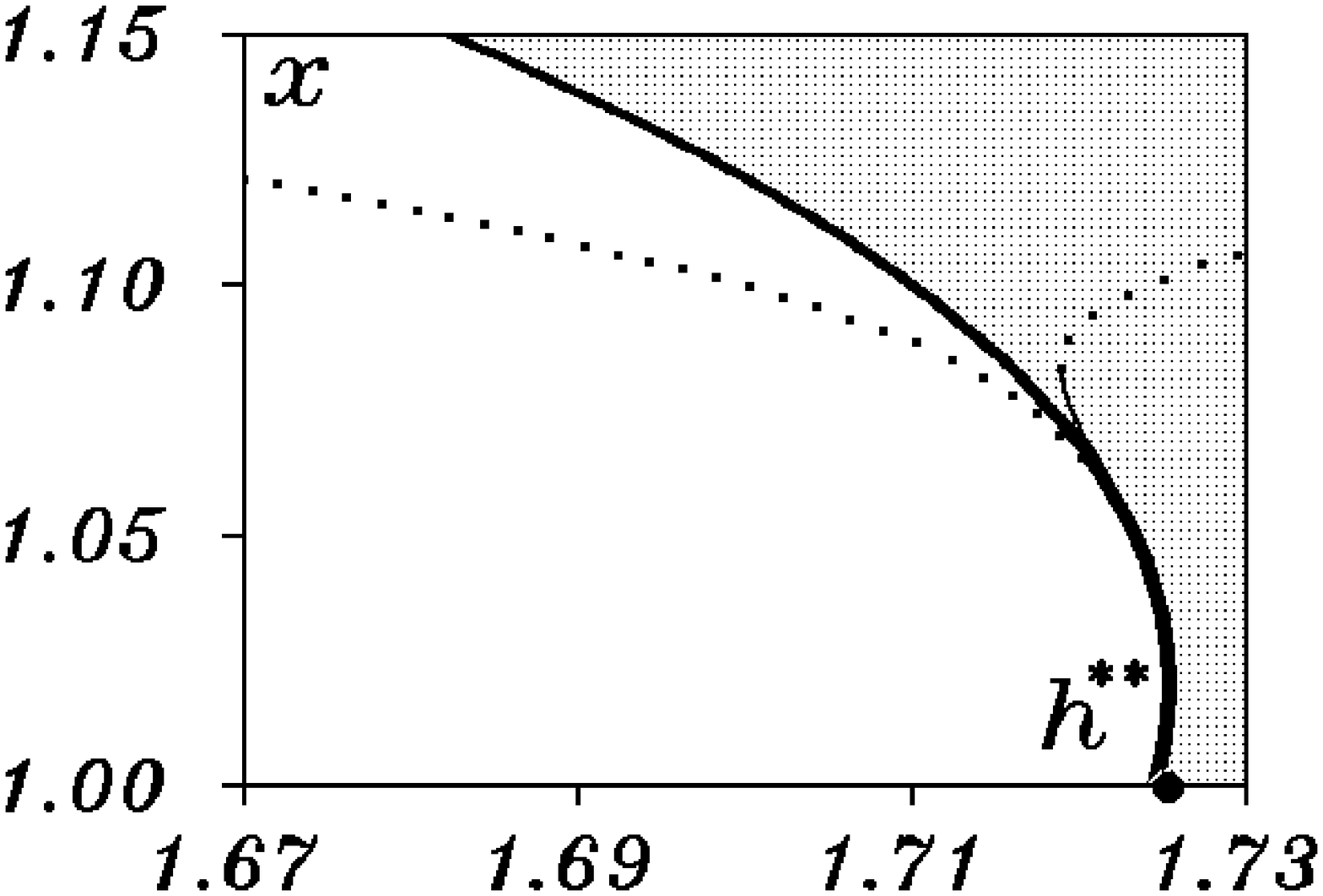}}
\end{picture}

\vspace*{3cm} Fig. 17. Bifurcation diagram in the case $e'=0.02 <
e'_*$ (a) and its fragments (b and c). For different values of
$h$ the diagram presents equilibrium solutions to system (6.1)
located on the $Ox$ axis. Thin solid and dashed lines correspond
to the families of stable and unstable solutions, respectively.
Thick line shown location of the points at which indeterminacy
curve $\Gamma(h)$ intersects axis $Ox$. Vertical segments of the
region marked in black are the segments of axis $Ox$ lying in the
forbidden area $M(h)$ at a corresponding value of $h$.

\newpage \pagestyle{empty}

\unitlength=1mm
\begin{picture}(0,30)
\put(20,-70){\includegraphics[width=10.0cm,height=8.0cm]{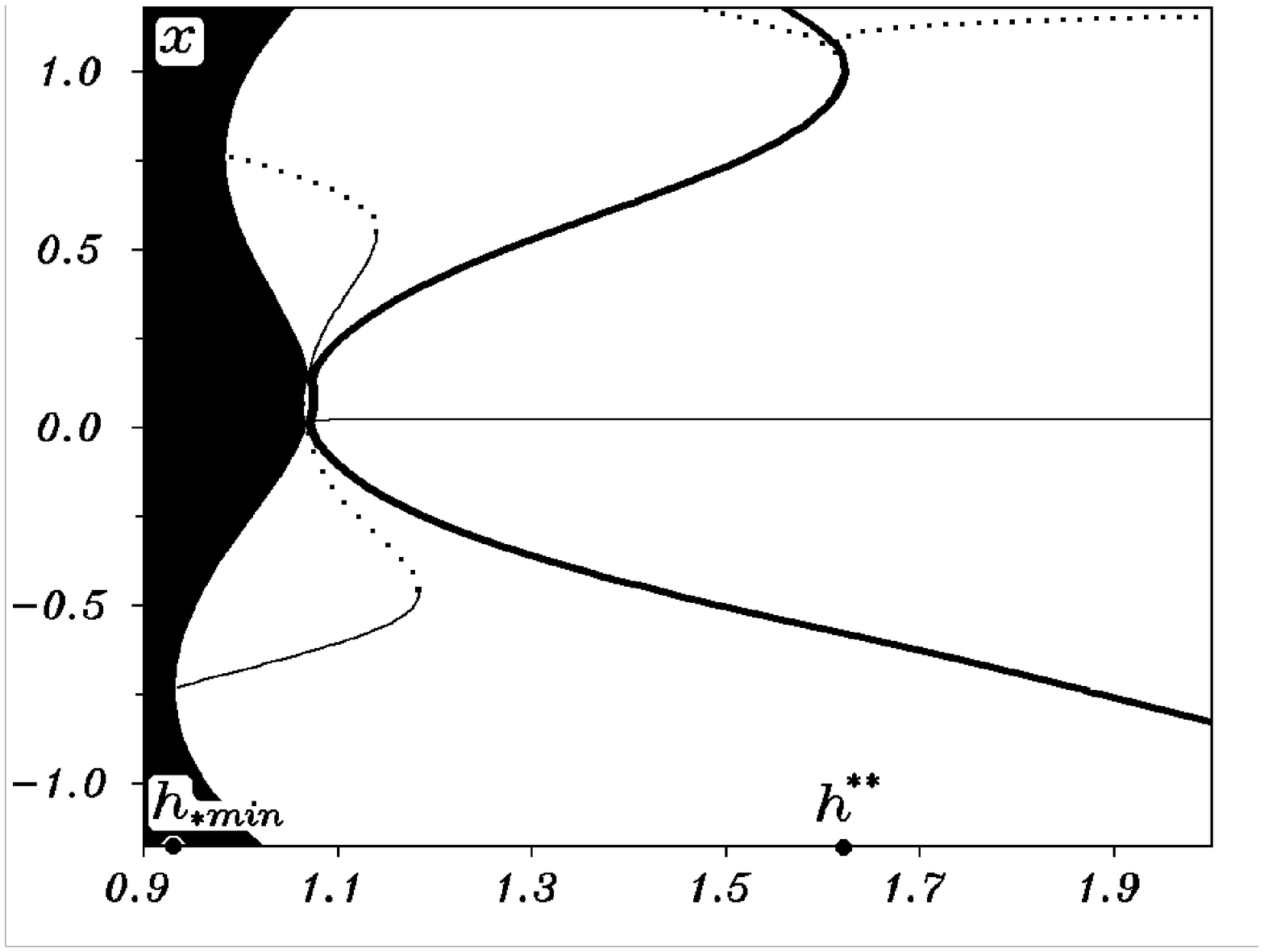}}
\end{picture}

\begin{picture}(0,150)
\put(40,10){\includegraphics[width=6.0cm,height=5.0cm]{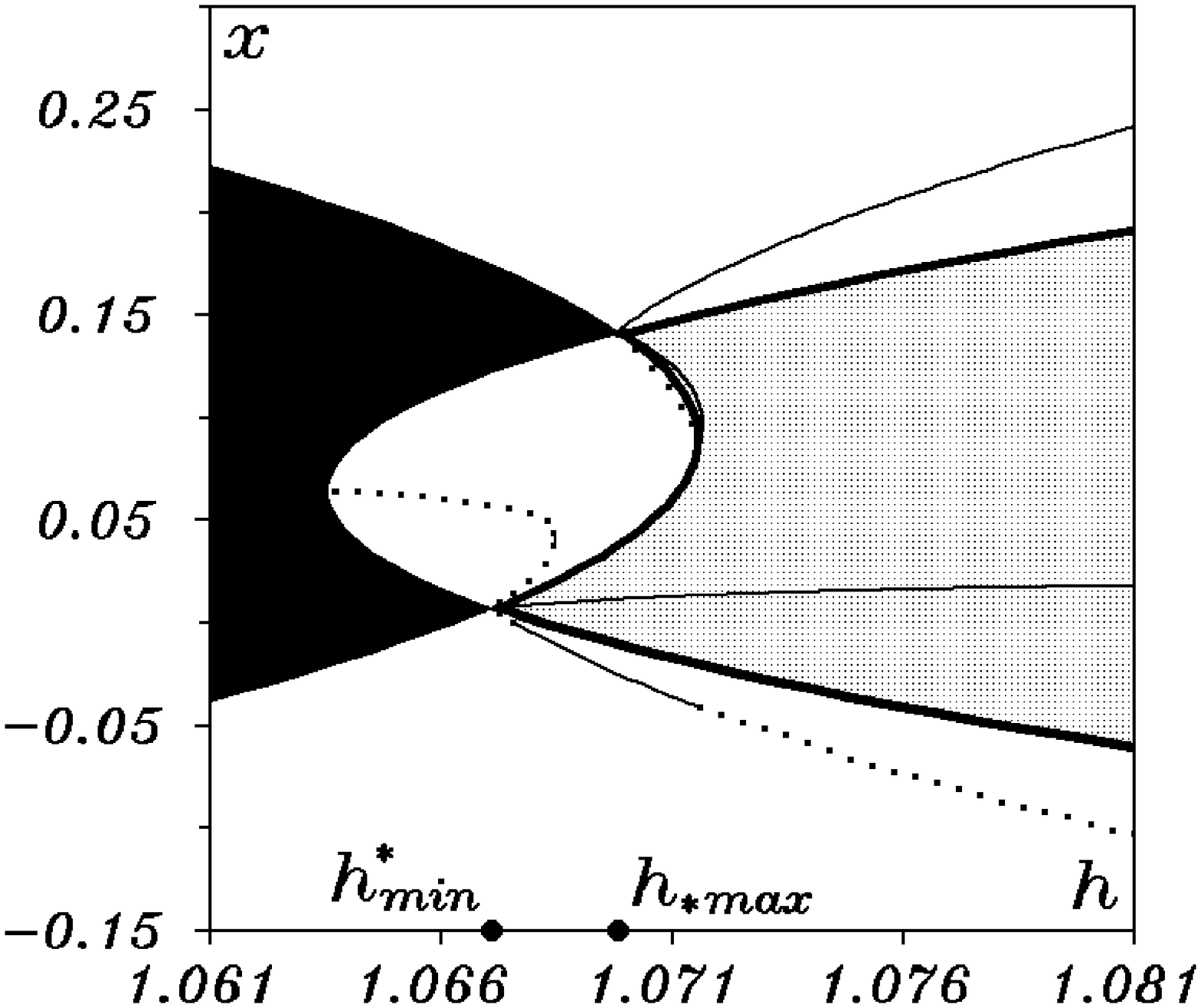}}
\end{picture}

\begin{center}
Fig. 18. Bifurcation diagram in the case $e'=0.048
> e'_*$ (a) and its enlarged fragment corresponding to variation
of $h$ in the interval $I_W$ (b).
\end{center}
\newpage \pagestyle{empty}
\unitlength=1mm
\begin{picture}(0,40)
\put(20,-90){\includegraphics[width=9.9cm,height=7.0cm]{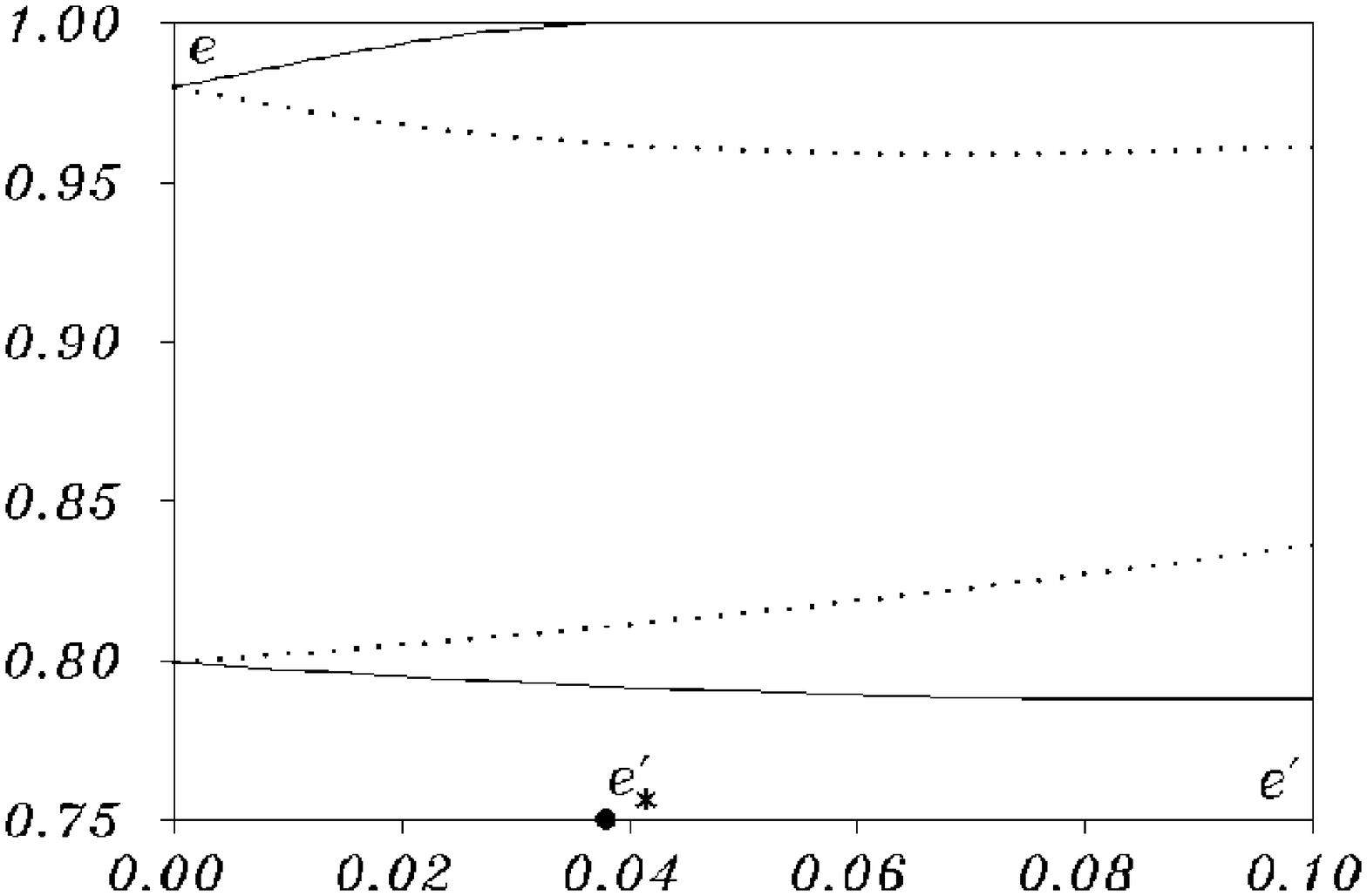}}
\end{picture}

\vspace*{10cm} Fig. 19. The value of eccentricity $e$ in periodic
solutions to the three-body problem corresponding to equilibrium
solutions to system (3.1). Solid and dashed lines correspond to
the families of stable and unstable solutions, respectively.

\newpage \pagestyle{empty}
\unitlength=1mm
\begin{picture}(0,70)
\put(30,-30){\includegraphics[width=9.0cm,height=9.0cm]{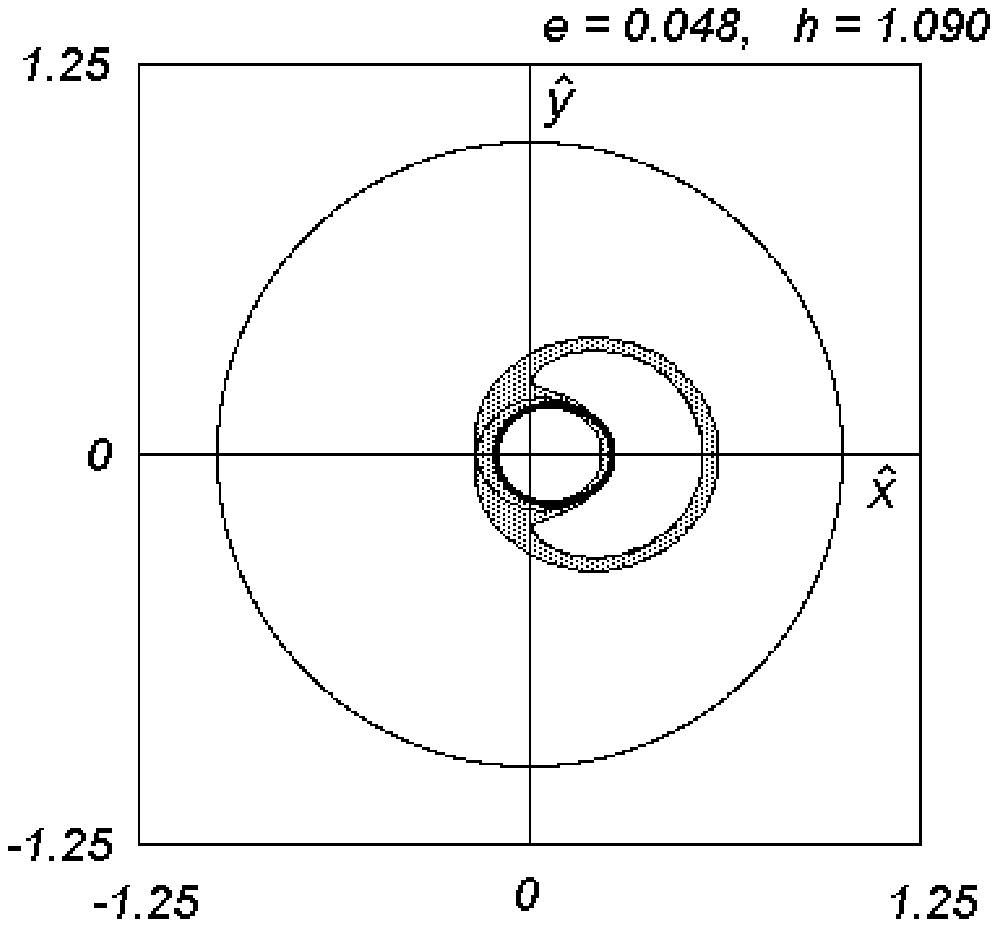}}
\end{picture}

\begin{picture}(0,90)
\put(30,-30){\includegraphics[width=9.0cm,height=9.0cm]{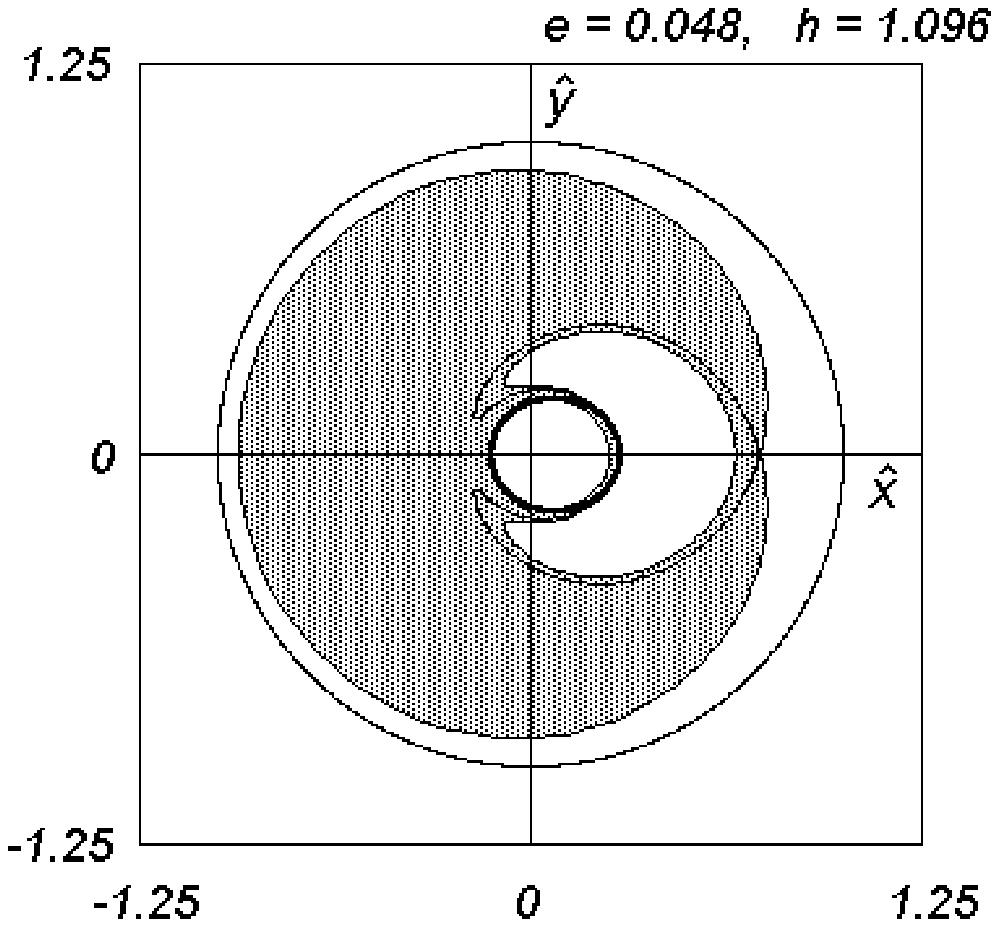}}
\end{picture}

\begin{center}
\vspace*{3cm} Fig. 20. The region of adiabatic chaos before (a)
and after (b) reconnection of separatrices.
\end{center}


\begin{thebibliography}{99}

\bibitem{hill}
Hill G.W. Illustrations of periodic solutions in the problem of
three bodies. {\it Astr. J.} (1902), {\bf 22}, 117--121.

\bibitem{sinclair}
Sinclair A.T. Periodic solutions close to commensurabilities in
the three body problem. {\it Mon. Not. Roy. Astr. Soc.} (1970),
{\bf 148}, 325--351.

\bibitem{scholl}
Scholl H., Froeschl\'e C. Asteroidal motion at the 3/1
commensurability. {\it Astron. Astrophys.} (1974), {\bf 33},
455--458.

\bibitem{w1}
Wisdom J. The origin of the Kirkwood gaps: a mapping for the
asteroidal motion near the 3/1 commensurability. {\it Astr. J.}
(1982), {\bf 87}, 577--593.

\bibitem{w2}
Wisdom J. Chaotic behavior and the origin of the 3/1 Kirkwood gap.
{\it ICARUS} (1983), {\bf 56}, 51--74.

\bibitem{w3}
Wisdom J. A perturbative treatment of motion near the 3/1
commensurability. {\it ICARUS} (1985), {\bf 63}, 272--286.

\bibitem{n}
7. Neishtadt, A.I., Sudden Changes of Adiabatic Invariant at
Crossing the Separatrix and Origin of the 3:1 Kirkwood Gap. {\it
Dokl. Akad. Nauk SSSR} (1987), {\bf 295}, 47--50.

\bibitem{koiller}
Koiller J., Balthazar J.M., Yokoyama T. Relaxation-chaos
phenomena in celestial mechanics. {\it Physica D} (1987), {\bf
26}, 85--122.

\bibitem{mg1}
Morbidelli A., Giorgilli A. On the dynamics in the asteroids
belt. Part I: general theory. {\it Celest. Mech.} (1990), {\bf
47}, 145--172.

\bibitem{mg2}
Morbidelli A., Giorgilli A. On the dynamics in the asteroids
belt. Part II: detailed study of the main resonances. {\it
Celest. Mech.} (1990), {\bf 47}, 173--204.

\bibitem{hc}
Henrard J., Caranicolas N.D. Motion near the 3/1 resonance of the
planar elliptic restricted three body problem. {\it Celest. Mech.
Dyn. Astron.} (1990), {\bf 47}, 99--121.

\bibitem{f}
Ferraz-Mello S., Klafke J.C. A model for the study of
very-high-eccentricity asteroidal motion. The 3:1 resonance. {\it
Predictability, Stability and Chaos in N-body Dynamical Systems.}
Ed. Roy A.E. New York: Plenum Press (1991), 177--184.

\bibitem{h1}
Hadjidemetriou J.D. The elliptic restricted problem at the 3:1
resonance. {\it Celest. Mech. Dyn. Astron.} (1992), {\bf 53},
151--183.

\bibitem{h2}
Hadjidemetriou J.D. Asteroid motion near the 3:1 resonance. {\it
Celest. Mech. Dyn. Astron.} (1993), {\bf 56}, 563--599.

\bibitem{grau}
Grau M. Critical point families at the 3:1 resonance. {\it
Celest. Mech. Dyn. Astron.} (1995), {\bf 61}, 389--401.

\bibitem{v}
Varadi F., Kaula W.K. Chaos in the 3:1 mean-motion resonance
revisited. {\it Planet. Space Sci.} (1999), {\bf 47}, 997--1003.

\bibitem{ns}
Neishtadt A.I., Sidorenko V.V. Wisdom system: dynamics in the
adiabatic approximation. {\it Celest. Mech. Dyn. Astron.} (2004),
{\bf 90}, 307--330.

\bibitem{vash1}
Vashkov'yak M.A. Method of Numerical Averaging in the Problem of
Evolution of Orbits of Resonance Asteroids. {\it Kosm. Issled.}
(1989), {\bf 27}, 9--14.

\bibitem{vash2}
Vashkov'yak M.A. Investigation of Evolution of Almost Circular
Orbits of Distant Satellites in Resonance 3:1 with the Moon Using
a Numerical-Analytical Method. {\it Kosm. Issled.} (1989), {\bf
27}, 817--826.

\bibitem{arnold}
Arnol'd V.I., Kozlov V.V., Neishtadt A.I. Matematicheskie aspekty
klassicheskoi i nebesnoi mekhaniki (Mathematical Aspects of
Classical and Celestial Mechanics). Moscow: Editorial URSS (2002).

\bibitem{nts}
Neishtadt A.I, Treschev D.V., Sidorenko V.V. Stable periodic
motions in the problem of passage through a separatrix. {\it
CHAOS} (1997), {\bf 7}, 2--11.

\bibitem{w4}
Tittemore W.C., Wisdom J. Tidal evolution of the uranian
satellites III. Evolution through the Miranda-Umbriel 3:1,
Miranda-Ariel 5:3, and Ariel-Umbriel 2:1 mean-motion
commensurabilities. {\it ICARUS} (1990), {\bf 85}, 394--443.

\bibitem{Kinoshita}
Ji J., Kinoshita H., Liu L., Li G. Could the 55 Cancri planetary
system really be in the 3:1 mean motion resonance. {\it Astrophys.
J.} (2003), {\bf 585}, L139--L142.
\end{thebibliography}
\end{document}